\documentclass[12pt]{article}
\usepackage{amsmath,amssymb, graphicx,euscript,url,bm}
\usepackage{physics}
\usepackage{bm}
\usepackage{mathrsfs}
\usepackage{mathtools}
\usepackage{cite}
\usepackage{siunitx}
\usepackage{float}
\usepackage[font=small,skip=0pt]{caption}
\usepackage[caption = false]{subfig}
\usepackage{slashed}
\def\weff{\omega_{\mbox{\scriptsize eff}}}

\def \pslash{\slashed{p}}

\begin{document}
\begin{titlepage}
\begin{flushright}
WSU-HEP-2402\\
May 8, 2024\\
\end{flushright}

\vspace{0.7cm}
\begin{center}
\Large\bf\boldmath
Quasielastic lepton-nucleus scattering and the correlated Fermi gas model
\unboldmath
\end{center}

\vspace{0.8cm}
\begin{center}
{\sc Bhubanjyoti Bhattacharya,$^{(a,b)}$ Sam Carey,$^{(b)}$ Erez O. Cohen,$^{(c)}$and Gil Paz$^{(b)}$ }\\
\vspace{0.4cm}
{\it 
$^{(a)}$ Department of Natural Sciences, \\Lawrence Technological University, 
Southfield, Michigan 48075, USA
}\\
\vspace{0.3cm}
{\it 
$^{(b)}$ Department of Physics and Astronomy, \\
Wayne State University, Detroit, Michigan 48201, USA}\\
\vspace{0.3cm} 
{\it 
$^{(c)}$ Physics Department, \\ Nuclear Research Center-Negev,\\ P.O. Box 9001, 84190 Beer-Sheva, Israel
}
\end{center}
\vspace{0.7cm}

\begin{abstract}
  \vspace{0.2cm}
  \noindent
The neutrino research program in the coming decades will require improved precision. A major source of uncertainty is the interaction of neutrinos with nuclei that serve as targets for such experiments. Broadly speaking, this interaction often depends, e.g., for charge-current quasielastic scattering, on the combination of ``nucleon physics," expressed by form factors, and ``nuclear physics," expressed by a nuclear model. It is important to get a good handle on both. We present a fully analytic implementation of the correlated Fermi gas model for electron-nucleus and charge-current quasielastic neutrino-nucleus scattering. The implementation is used to compare separately form factors and nuclear model effects for both electron-carbon and neutrino-carbon scattering data.
\end{abstract}
\vfil

\end{titlepage}

\newpage

\section{INTRODUCTION}

Current and upcoming experiments involving lepton-nucleon scattering aim to precisely measure parameters in the Standard Model Lagrangian that describe leptonic interactions as well as to uncover neutrino nonstandard interactions (NSI). Such precision measurements require a better control of systematic uncertainties in lepton-nucleus interactions. The cross section for a charged lepton or a charge-current quasielastic (CCQE) neutrino scattering off a nucleus is determined by folding the lepton-quark interaction twice. Each folding gives rise to a source of systematic uncertainty. The first involves nucleon form factors needed to fold the lepton-quark interaction into the lepton-nucleon interaction. The result is the scattering cross section on a single nucleon. The second folding is needed in going from the nucleon to the nuclear level, which involves a nuclear model and results in the cross section on a nucleus consisting of multiple nucleons. Ideally, we would like to have separately a good control for each source of uncertainty.

It can be argued that the majority of the research in this field has focused on the nuclear level, where the nucleon level has received less study. This seems plausible for uncertainties arising from the electromagnetic form factors, since they can be extracted from electron-nucleon scattering ; see, for example, the parametrizations in Refs. \cite{Bradford:2006yz} and \cite{Borah:2020gte}. Uncertainties from the axial form factor are much harder to control.  Historically, a dipole form factor was assumed for the axial form factor. For example, the comparative study \cite{Sobczyk:2017mts}, based on Ref. \cite{Sobczyk:2019yiy}, compared six nuclear models: Benhar's spectral function with and without the final-state interactions \cite{Benhar:1994hw,Benhar:2010nx,Petraki:2002nb,Ankowski:2014yfa}, the Valencia spectral function \cite{Gil:1997bm,Nieves:2004wx,Nieves:2017lij,FernandezdeCordoba:1991wf}, the Giessen Boltzmann-Uehling-Uhlenbeck (GiBUU) model \cite{Buss:2011mx,Gallmeister:2016dnq}, and the local and global Fermi gas models, all using the dipole model for the axial form factor \cite{Sobczyk:2019yiy}. In addition, various quantum many-body methods were used to evaluate the inclusive cross section. These include Green's function Monte Carlo (GFMC) method, coupled-cluster approach, short-time approximation, mean-field approach using the relativistic plane-wave impulse approximation (RPWIA), and the Super-Scaling Approach (SuSA). See Sec. \MakeUppercase{\romannumeral 5}.D. in the white paper \cite{Ruso:2022} and references within.

The dipole model is not motivated from first principles and its usage can underestimate the uncertainties.  This issue was highlighted in the the analysis done by the MiniBooNE Experiment \cite{MiniBooNE:2010bsu} that seemed to require a higher axial mass than the preceding world-average value to explain the data.  In Refs. \cite{Bhattacharya:2011ah, Bhattacharya:2015mpa} a $z$-expansion based parametrization was introduced for the axial form factor. Assuming a definite nuclear model, namely, the relativistic  Fermi gas (RFG) model \cite{Moniz:1971mt}, an axial mass was extracted from the MiniBooNE data using $z$-expansion based parametrization that was consistent with the preceding world-average values. 

In the RFG model a nucleon within a nucleus is assumed to occupy momentum states only up to the Fermi momentum. Data from the last two decades have shown that this is not the case. Inclusive measurements involving large momentum transfer ($Q^2 > 1.5\,$ GeV$^2$), showed that approximately $20\%$ of nucleons have momentum greater than the Fermi momentum \cite{CLAS:2003eih, CLAS:2005ola, Fomin:2011ng}. Almost all high-momentum nucleons appear in short-range correlated (SRC) pairs, predominantly neutron-proton ones, and contribute most of the kinetic energy carried by nucleons in nuclei. Exclusive measurements on $^{12}$C and $^4$He have led to direct observations of such SRC pairs \cite{E850:2000sud, Piasetzky:2006ai, JeffersonLabHallA:2007lly, Subedi:2008zz, LabHallA:2014wqo}.  An extension of the RFG model that takes into account the high momentum tail beyond the nuclear Fermi momentum due to SRC pairing is the correlated Fermi gas (CFG) model suggested in Ref. \cite{Hen:2014yfa}. In the CFG model, the nucleon momentum distribution is constant below the Fermi momentum and has a small high-momentum tail above the Fermi momentum.

The CFG model has been used to study the equation of state of nucleonic stars and its effects on stellar properties such as maximum mass and particle fraction\cite{Cai:2022grw, Raithel:2021hye, Raithel:2019gws, Hen:2016ysx}. The CFG model has also been implemented to study the EMC (European Muon Collaboration) Effect, where the quark-gluon structure of a nucleon bound in an atomic nucleus is modified by the surrounding nucleons \cite{CLAS:2019vsb}. One goal of this paper is to present a fully analytic implementation of the CFG model for electron-nucleus and CCQE neutrino-nucleus scattering. \footnote{See Ref. \cite{GENIE:2021npt} for a related implementation in GENIE v3.2\,.}  

A second goal of this paper is to use this implementation to compare \emph{separately} form factors and nuclear model effects for both electron-carbon and neutrino-carbon scattering data. Lack of such a control is both a long-standing issue and a current topic. The first arises from the fact that many explanations for the MiniBooNE data \cite{MiniBooNE:2010bsu} assumed  the dipole axial form factor model and only tried to modify the nuclear effects. These include the addition of multinucleon processes (2p2h contributions with or without $\Delta$ excitations) 
\cite{Martini:2009, Martini:2010, Benhar:2010, Barbaro:2011, Amaro:2011, Nieves:2011, Nieves:2011b, Bodek:2011, Meucci:2011}. The studies of Refs. \cite{Bhattacharya:2011ah, Bhattacharya:2015mpa} fixed the nuclear model and allowed for a flexible axial form factor. The lack of control is a current topic, since in the last few years, many extractions of the axial from factors from experimental data \cite{Meyer:2016oeg, MINERvA:2023avz} and lattice QCD \cite{RQCD:2019jai,Park:2021ypf,Djukanovic:2022wru, Jang:2023zts, Alexandrou:2023qbg} became available. A better control on the axial form factor will potentially allow to separate the nucleon and nuclear effects.

The structure of the paper is as follows: in Sec. \ref{sec:Models}, we review the relation between the cross section and the nuclear tensor, and present the construction of the nuclear tensor using the nucleon form factors for the RFG and CFG models. In Sec. \ref{sec:CFG}, we present the analytic implementation of the CFG model for lepton-nucleus scattering. We compare the CFG predictions to electron-carbon data in Sec. \ref{sec:electron} and to neutrino MiniBooNE data in Sec. \ref{sec:neutrino}. For both we consider separately form factor effects by comparing different form factor parametrizations,  and  nuclear effects by comparing  the RFG and CFG models. We summarize our findings in Sec. \ref{sec:Summary}. Some more technical details of the paper are relegated to the appendixes.

\section{MODELS}\label{sec:Models}
\subsection{Lepton-nucleus cross section}
We study processes in which an incoming lepton, neutral or charged, with four-momentum $k$ that scatters to a lepton with four-momentum $k^\prime$ off a target with four-momentum $p^T$. The differential cross section is expressed in terms of the nuclear tensor $W_{\mu\nu}$ defined below. Considering the possibility of both vector and axial currents, we can decompose  $W_{\mu\nu}$ as a linear combination of scalar functions $W_i$ as \cite{Bhattacharya:2011ah}
\begin{equation}
W_{\mu\nu}=-g_{\mu\nu}W_1+\frac{p_\mu^Tp_\nu^T}{m_T^2}W_2-\frac{i\epsilon_{\mu\nu\rho\sigma}p_T^\rho q^\sigma}{2m_T^2}W_3+\frac{q_\mu q_\nu}{m_T^2}W_4+\frac{p_\mu^Tq_\nu+q_\mu p_\nu^T}{2m_T^2}W_5\,,
\end{equation}
where $q=k-k^\prime$ and $m_T$ is the target mass. Defining $E_\ell\equiv{k^\prime}^0$ and $\vec{P}_\ell\equiv\vec{k}^\prime$ the energy and momentum of the final-state lepton, the (anti)neutrino-nucleus cross section is \cite{Bhattacharya:2011ah}
\begin{multline}\label{eq:xs_neutrino}
\frac{d\sigma^{\,\nu}_{\rm nuclear}}{dE_\ell\, d\cos\theta_\ell}=
\frac{G_F^2 |\vec{P}_\ell |}{16 \pi^2\, m_T }
\Bigg\{2 (E_\ell -|\vec{P}_\ell |\cos \theta_\ell)\,W_1
+( E_\ell  + |\vec{P}_\ell |\cos \theta_\ell) W_2
\\
\pm \frac{1}{m_T}\Big[(E_\ell -|\vec{P}_\ell |\cos \theta_\ell)(E_\nu+E_\ell )-m^2_\ell \Big]W_3
+ \frac{m_\ell ^2}{m_T^2}(E_\ell -|\vec{P}_\ell |\cos \theta_\ell) W_4- \frac{m_\ell ^2}{m_T}\, W_5\Bigg\}\,,
\end{multline}
where the upper (lower) sign is for neutrino (antineutrino) scattering. 

Neglecting the electron mass, the electron-nucleus cross section is  
\begin{equation}\label{eq:xs_electron}
\frac{d\sigma^{\,e}_{\rm nuclear}}{dE_\ell\, d\cos\theta_\ell}=\frac{\alpha^2 E_\ell^2}{2q^4\, m_T}\Big[2W_1(1-\cos \theta_\ell)+W_2(1+\cos \theta_\ell)\Big].
\end{equation}
The nuclear tensor $W_{\mu\nu}$ is formally related to matrix elements of the vector and axial current between the initial and final nuclear states.  We can relate it to the single nucleon tensor, $H_{\mu\nu}$ for both the RFG and CFG models by using a ``statistical" approach presented explicitly in Ref. \cite{Bhattacharya:2011ah}. Denote by $n_i(\bm{p})$ the momentum distribution of the initial nucleon momentum $\bm p$. The final-state nucleon momentum $\bm p^\prime$  phase space is limited by a factor of $\left[1-n_f(\bm{p^\prime})\right]$ from the Fermi-Dirac statistics.  The nuclear tensor $W_{\mu\nu}$ can be expressed in terms of the nucleon tensor $H_{\mu\nu}$, as follows: 
\begin{equation}\label{eq:General_W_H_relation}
W_{\mu\nu}=\int \frac{d^3p}{(2\pi)^3}\, \frac{m_T}{E_p}\,2V n_i(\bm{p}) \int \frac{d^3p'}{(2\pi)^3 2E_{p^\prime}}(2\pi)^4\delta^4(p-p^\prime +q)H_{\mu\nu}\left[1-n_f(\bm{p^\prime})\right]\,;
\end{equation} 
see Ref. \cite{Bhattacharya:2011ah} for details. Accounting for two possible spin states, the number of nucleons $N$ of a certain type, namely, protons or neutrons, determines  the normalization factor $V$ as
\begin{equation} \label{eq:normalization}
N=\int \frac{d^3p}{(2\pi)^3}\, 2V n_i(\bm{p}). 
\end{equation}
In the following, we use $N=A/2$, where $A$ is the total number of nucleons.

Since we will not consider the kinematics of the final-state nucleon in this paper, we can integrate over $d^3p'$ using the three-momentum delta function. We also incorporate a binding energy, $\epsilon_b$, 
by the replacements
\begin{equation}\label{eq:replacements}
p^0 \to \epsilon_{\bm{p}} \equiv E_{\bm p} - \epsilon_b \,,
\quad
p^{\prime 0} \to \epsilon^\prime_{\bm{p}^\prime}  \equiv E_{\bm{p}^\prime} \,, 
\end{equation}
where $E_{\bm{p}} \equiv \sqrt{m_N^2 + |\bm{p}|^2}$.  The resulting relation between $W_{\mu\nu}$ and $H_{\mu\nu}$ is \cite{Bhattacharya:2011ah}
\begin{eqnarray} \label{eq:convolution}
W_{\mu\nu} \equiv&\int d^3{p}\,f({\bm p},q^0,\bm{q})
H_{\mu\nu}(\epsilon_{\bm{p}}, \bm{p}; q^0, \bm{q} )  \,,
\end{eqnarray}
with 
\begin{equation}\label{eq:f_definition}
f({\bm p},q^0,\bm{q})
=
\frac{m_T V}{4\pi^2}
 \, n_i(\bm{p}) 
[1-n_f(\bm{p}+\bm{q})] 
\frac{\delta( \epsilon_{\bm{p}} - \epsilon^\prime_{\bm{p}+\bm{q}} + q^0 )}{\epsilon_{\bm{p}} \epsilon^\prime_{\bm{p}+\bm{q}}} \,. 
\end{equation}
We discuss $f({\bm p},q^0,\bm{q})$ for the RFG and CFG cases below.   The nucleon tensor $H_{\mu\nu}$ is given by 
\begin{equation}
H_{\mu\nu} = 
{\rm Tr}[  ( \pslash^\prime + m_{p^\prime} )\Gamma_\mu(q) (\pslash+m_p )\bar{\Gamma}_\nu(q) ] \,,
\end{equation}
where $\Gamma_\mu(q)$ is defined via the matrix element of the electromagnetic or charged weak current $J_\mu$ as
\begin{align}
\langle p^\prime|J_\mu|p\rangle&=
\bar u(p^\prime) \Gamma_\mu(q) u(p)\,.
\end{align}
$\Gamma_\mu(q)$ can be expressed in term of form factors: 
\begin{equation}\label{eq:vertex}
 \Gamma_\mu(q) 
= \gamma_\mu F_1(q^2) + \frac{i}{2m_N}\sigma_{\mu\nu}q^\nu F_2(q^2)
+ \gamma_{\mu}\gamma_5 F_A(q^2)+\frac{q_\mu} {m_N} \gamma_5 F_P(q^2)\,,
\end{equation}
where we assume time-reversal invariance. For neutrino scattering we also assume isospin symmetry that allows us to relate the neutron-to-proton $F_{1,2}$ to the electromagnetic form factors $F_{1,2}$ of the proton and neutron. For charged-lepton scattering, the nucleon mass is replaced by the mass of the proton ($m_p$) or neutron ($m_n$). For neutrino scattering, we assume isospin symmetry and take $m_N=m_p=m_n$. Note that the inclusion of the binding energy in Eq. (\ref{eq:replacements}) leads to the nonconservation of the vector current. This can be fixed by including an extra term in Eq. (\ref{eq:vertex}) as suggested in Refs. \cite{Smith:1972xh, Forest:1983ahx}. We ignore this in the following, as it only leads to a correction to the cross section of the order $m_l^2$ \cite{Smith:1972xh}.  Based on these symmetries, $H_{\mu\nu}$ can be decomposed as  
\begin{equation}
H_{\mu\nu} = -g_{\mu\nu} H_1 + \frac{p_\mu p_\nu}{m_N^2} H_2 - i\frac{ \epsilon_{\mu\nu\rho\sigma}}{2 m_N^2} p^\rho q^\sigma
H_3 + \frac{q_\mu q_\nu}{m_N^2} H_4  
+ \frac{(p_\mu q_\nu + q_\mu p_\nu)}{2m_N^2 } H_5
\,. 
\end{equation}
The $H_i$'s are expressed in terms of the form factors $F_i$ as \cite{Bhattacharya:2011ah}
\begin{eqnarray}\label{eq:H_i}
H_1 &=& 8 m_N^2 F_A^2 -2q^2 \left[ (F_1 + F_2)^2 + F_A^2 \right]\,\nonumber\\   
H_2 &=& H_5 =  8 m_N^2 \left( F_1^2 + F_A^2\right) -2q^2 F_2^2\,,\nonumber\\  
H_3 &=& -16m_N^2\, F_A (F_1 + F_2 )\,,\nonumber\\  
H_4 &=& -\frac{q^2}{2}\left(F_2^2 + 4F_P^2 \right)-2m_N^2 F_2^2 - 4m_N^2 \left( F_1F_2 + 2 F_A F_P\right) \,.
\end{eqnarray}

Combining these expressions, the cross section is expressed by the nuclear tensor $W_{\mu\nu}$ that is a convolution of the single nucleon tensor $H_{\mu\nu}$ and a nuclear model parametrized by the initial and final nucleon momentum distributions.

We now discuss the nucleon momentum distributions for the RFG and the CFG models. 

\subsection{Relativistic Fermi gas model}
In the RFG model the distributions of neutrons and protons are 
\begin{equation} \label{niRFG}
n_i(\bm{p}) = \theta( p_F - |\bm{p}| ) \,,
\quad 
n_f(\bm{p}^\prime) = \theta( p_F - |\bm{p}^\prime|) \,,
\end{equation}
where $p_F$ is a parameter of the model.  Using Eq. (\ref{eq:normalization}), we find that the normalization factor $V$ is 
\begin{equation}
\frac{A}{2}  =  \int\frac{d^3 p}{(2\pi)^3} 2V\,n_i(\bm{p})  \implies V = \frac{A}2\frac{3\pi^2}{p_F^3} \,.
\end{equation}
From these, explicit expressions can be derived for $W_i$. In particular, 
\begin{align} \label{Wi}
W_1 &= a_1 H_1 + \frac12( a_2 - a_3) H_2 \,,\nonumber\\
W_2 &= \left[ a_4 + \frac{\omega^2}{|\bm{q}|^2}a_3- 2\frac{\omega}{|\bm{q}|} a_5 + \frac12\left(1-\frac{\omega^2}{|\bm{q}|^2}\right) (a_2 - a_3) \right] H_2 \,,\nonumber\\
W_3 &= \frac{m_T}{m_N} \left( a_7 - \frac{\omega}{|\bm{q}|} a_6 \right) H_3 \,,\nonumber\\
W_4 &= \frac{m_T^2}{m_N^2} \left[ a_1 H_4 + \frac{m_N}{|\bm{q}|} a_6 H_5 + \frac{m_N^2}{2 |\bm{q}|^2} (3a_3 - a_2) H_2 \right] \,,\nonumber\\
W_5 &= \frac{m_T}{m_N} \left( a_7 - \frac{\omega}{|\bm{q}|} a_6 \right) H_5 + \frac{m_T}{|\bm{q}|}\left[ 2 a_5 + \frac{\omega}{|\bm{q}|}(a_2 - 3 a_3) \right] H_2 \,,
\end{align}
where $\omega=q^0$ and 
\begin{align}
a_1 &= \int d^3 {p} \, f({\bm p},q)\,, 
& a_2 = \int d^3 {p} \, f({\bm p},q)\,\frac{|\bm{p}|^2}{m_N^2}\,,\nonumber\\
a_3 &= \int d^3 {p} \, f({\bm p},q)\, \frac{(p^z)^2}{m_N^2} \,, 
& a_4 = \int d^3 {p} \, f({\bm p},q)\, \frac{\epsilon_{\bm{p}}^2}{m_N^2} \,,\nonumber\\
a_5 &= \int d^3 {p} \, f({\bm p},q)\, \frac{\epsilon_{\bm{p}} p^z}{m_N^2} \,, 
&a_6 = \int d^3 {p} \, f({\bm p},q)\, \frac{p^z}{m_N}\,,\nonumber\\
a_7 &=\int d^3 {p} \, f({\bm p},q)\, \frac{\epsilon_{\bm{p}}}{m_N} \,;
\end{align} 
see, for example, Ref. \cite{Bhattacharya:2011ah}. For the RFG model, the $a_i$ functions can be expressed in terms of three master integrals $b_i$ over the initial nucleon energy. The limits of these master integrals are determined by conservation of three-momentum and energy. We will encounter similar features in the CFG model.
\subsection{Correlated Fermi gas model}
We follow the model of Ref. \cite{Hen:2014yfa} with $\rho=\rho_0$ and change the momentum variable from $k$ to $p$. The momentum distribution there is given by
\begin{equation}
n^{SRC}_{SNM} (\bm{p})=\begin{cases} 
     A_0 & |\bm{p}|\leq p_F \\
      c_0 p_F/p^4& p_F\leq |\bm{p}|\leq \lambda p_F \\
      0 & |\bm{p}|\geq \lambda p_F,
   \end{cases}
\end{equation}
where $c_0=4.16\pm0.95$ and $\lambda\approx2.75\pm0.25$. For $\rho=\rho_0$, $A_0$ is given by  
\begin{equation}
A_0=\dfrac{3\pi^2}{p_F^3}\left[1-\left(1-\dfrac1{\lambda}\right)\dfrac{c_0}{\pi^2}\right],
\end{equation}
and determined by the normalization 
\begin{equation} 
1  = 2 \int\frac{d^3 p}{(2\pi)^3} n^{SRC}_{SNM} (\bm{p}).
\end{equation}
To obtain $n$ for the CFG model, we change the normalization to $A/2$. Thus, we define $n_{\mbox{\scriptsize CFG}} (\bm{p})$ to be \begin{equation}\label{eq:nCFG}
n_{\mbox{\scriptsize CFG}} (\bm{p})=\begin{cases} 
     1-\left(1-\dfrac1{\lambda}\right)\dfrac{c_0}{\pi^2}\equiv \alpha_0 & |\bm{p}|\leq p_F \\
      \dfrac{c_0}{3\pi^2} \left(\dfrac{p_F}{\bm{p}}
      \right)^{\!\!4} \equiv  \dfrac{\alpha_1}{|\bm{p}|^4}& p_F\leq |\bm{p}|\leq \lambda p_F \\
      0 & |\bm{p}|\geq \lambda p_F\,. 
   \end{cases}
\end{equation}
Taking $V=\left(A/2\right)3\pi^2 / p_F^3$,  $n_{\mbox{\scriptsize CFG}} (\bm{p})$ satisfies 
\begin{equation}
\frac{A}{2}  =  \int\frac{d^3 p}{(2\pi)^3} 2V\,n_{\mbox{\scriptsize CFG}} (\bm{p})\,.
\end{equation}

\begin{figure}[H]
\includegraphics[scale=0.5]{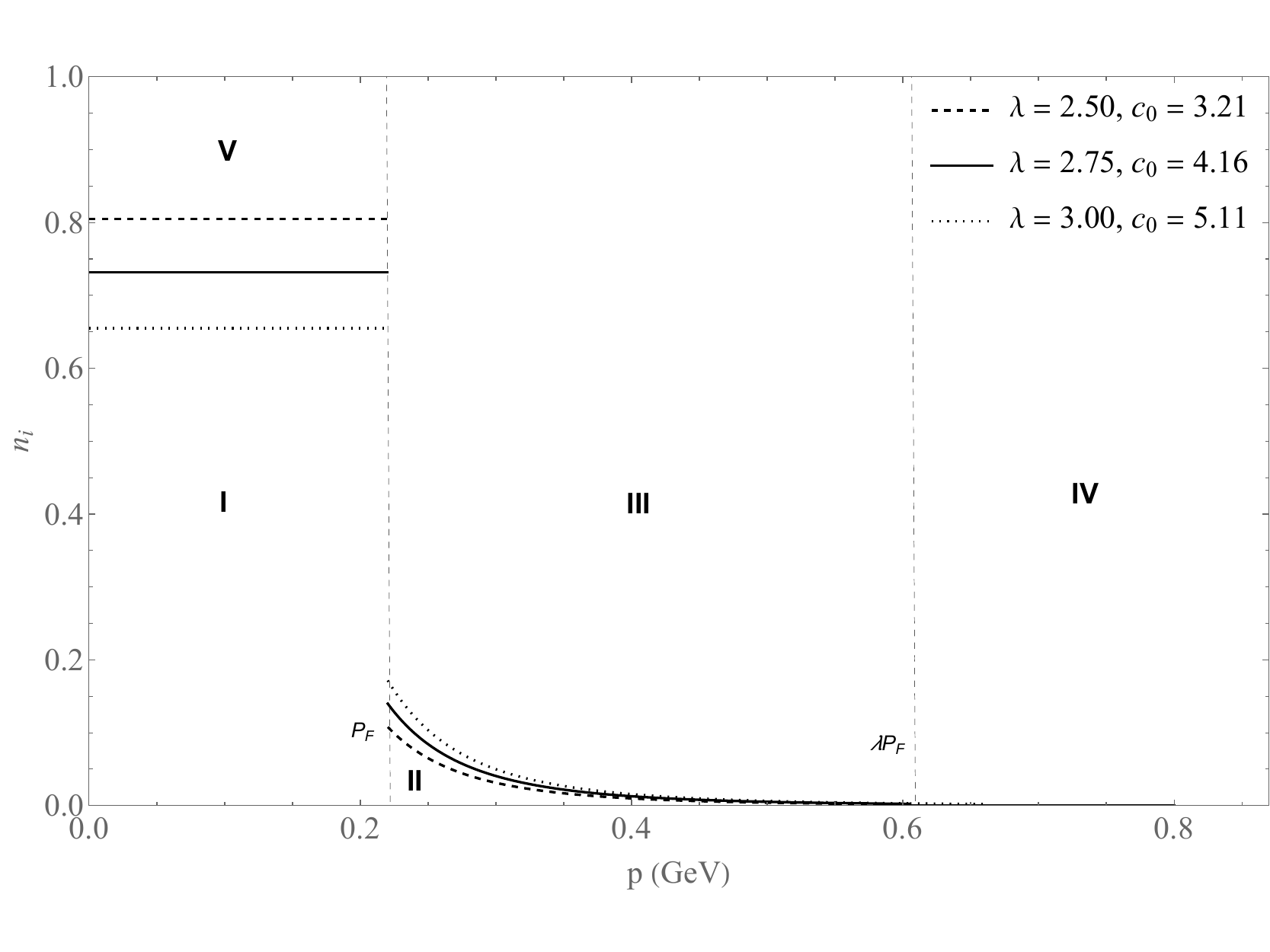}
\caption{\label{Fig:np} $n_{\mbox{\scriptsize CFG}} (\bm{p})$ as a a function of the magnitude of the three-momentum. The different lines correspond to extremal variation of $\lambda$ and $c_0$. The labels for the regions I--V are explained in the text.}
\end{figure}

$n_{\mbox{\scriptsize CFG}} (\bm{p})$ is plotted in Fig. \ref{Fig:np}. It exhibits a depleted Fermi gas region and a correlated high-momentum tail. The dashed and dotted lines correspond to the maximum and minimum limits for the high-momentum cut-off $\lambda$ and parameter $c_0$, with the central value represented by a straight line. Unlike the RFG model, where the nucleon can be either ``inside" or ``outside"  the nucleus, for the CFG model we can distinguish five regions. The initial nucleon can be in region I if $0\leq|\bm{p}|\leq p_F$, or in region II if $p_F\leq |\bm{p}|\leq \lambda p_F$. The final nucleon can be in three possible regions. Region III corresponds to $p_F\leq |\bm{p^\prime}|\leq \lambda p_F$, region IV to $\lambda p_F\leq |\bm{p^\prime}|$, and region V to $0\leq|\bm{p^\prime}|\leq p_F$. In the formal limit $\lambda\to 1$, regions II, III, and V vanish and we are left with the RFG model.  Since $\lambda\approx2.75\pm0.25$, this limit is never obtained in practice.

\section{IMPLEMENTATION OF THE CFG MODEL}\label{sec:CFG}
Using the properly normalized expression for $n_{\mbox{\scriptsize CFG}} (\bm{p})$ in Eqs. (\ref{eq:nCFG}) and (\ref{eq:convolution}), we derive explicit expressions for $W_i$. These will be more complicated than the ones for the RFG model since the momentum distribution is more complicated. Here we outline the implementation with detailed expressions relegated to the appendixes.

\subsection{Calculation of the nuclear tensor}

As discussed above, the initial nucleon can be in regions I or II, while the final nucleon can be in regions III, IV, or V.  Thus we have six possible transitions:  I$\to$ III, I$\to$ IV, I$\to$ V,  II$\to$ III, II$\to$ IV, and II$\to$ V.  For each of these transitions, we calculate $W_i$ separately and find the cross section. The total cross section will be the sum of these possible cases.  Since the momentum dependence in $n_{\mbox{\scriptsize CFG}}$ is either a constant or inversely proportional to the fourth power of the three-momentum, these six transitions depend on four possible combinations of initial and final momentum dependencies. Thus the five $W_i$ depend on the five $H_i$ via the seven functions $a_j$ as in the RFG model, but we need to define four sets of such functions. Each set in turn depends on three master $b$ integrals over the initial energy $E_{\bm{p}}$, analogous to the ones in the RFG case. The detailed expressions appear in Appendix \ref{app:ab_functions}.

To give a flavor of these expressions, assume that for a given value $\bm{p}$, $\bm{q}$ and $\omega$, there is a transition from region II$\to$ III. The energy-conserving delta function in Eq. (\ref{eq:f_definition}) requires that  $\epsilon_{\bm{p}} - \epsilon^\prime_{\bm{p}+\bm{q}} + q^0=0$ or $E_{\bm{p}+ \bm{q}} = E_{\bf p} +\omega_{\mbox{\scriptsize eff}}$, where $\omega_{\mbox{\scriptsize eff}}=\omega-\epsilon_b$. For the initial state this implies $|\bm p|^{-4}=\left(E_{{\bf p}}^2 - m^2_N\right)^{-2}$. Similarly, for the final state it implies, $|\bm p^\prime|^{-4}=|\bm p+\bm q|^{-4}=(E_{{\bf p} + {\bf q}}^2-m^2_N)^{-2}=\left[(E_{\bf p} +\omega_{\mbox{\scriptsize eff}})^2-m_N^2\right]^{-2}$. Thus the integrals over the function $f({\bm p},q^0,\bm{q})$ contain $\alpha_1\left(E_{{\bf p}}^2 - m^2_N\right)^{-2} \left(1 - \alpha_1 \left[(E_{\bf p} +\omega_{\mbox{\scriptsize eff}})^2-m_N^2\right]^{-2}\right)$. The energy-conserving delta function is used to fix the angle between $\bm{p}$ and $\bm{q}$ and the integral over $d^3p$ can be replaced by an integral over $E_p$. Similar considerations can be applied to other transitions. The explicit expressions are listed in Appendix \ref{app:ab_functions}. The limits of integration are different for each of the six transitions, and we discuss them next.

\subsection{Limits of integration}\label{sec:limits}

The definite integrals over  $E_{\bm{p}}$ have the limits $E_{\mbox{\scriptsize low}}\leq E_{\bm{p}}\leq E_{\mbox{\scriptsize high}}$. The values of $E_{\mbox{\scriptsize low}}$ and $E_{\mbox{\scriptsize high}}$ are different for each of the six possible transitions. 

Define $\omega=q^0, \weff = \omega - \epsilon_b, c= -\weff/ |\bm{q}|$,  $d= -(\weff^2 - |\bm{q}|^2)/(2|\bm{q}|m_N)$, and $E_F^\lambda=\sqrt{m_N^2+\left(\lambda p_F\right)^2}$. One condition arises from the constraints on the scattering angle. The scattering angle $\theta_{\bm{pq}}$ satisfies $-1\leq\cos \theta_{\bm{pq}}\leq 1$, or  
\begin{equation}
-1\leq\cos \theta_{\bm{pq}}\leq 1 \quad \Rightarrow \quad
-1 \le \frac{ \omega_{\rm eff}^2 -|\bm{q}|^2 + 2\omega_{\rm eff} E_{\bm{p}}}{ 
 2 |\bm{q}|\sqrt{E_{\bm{p}}^2-m_N^2} } \le 1. 
\end{equation}
The latter condition can be expressed as 
\begin{equation}
\left(\frac{E_{\bm{p}}}{m_N} - \frac{cd+\sqrt{1-c^2+d^2}}{1-c^2} \right)
\left(\frac{E_{\bm{p}}}{m_N} - \frac{cd-\sqrt{1-c^2+d^2}}{1-c^2} \right) \ge 0 \,.
\end{equation}
Defining $\Delta \equiv m_N(cd+\sqrt{1-c^2+d^2})/(1-c^2)$, this implies $\Delta \leq E_{\bm{p}}$. This condition holds for all regions.

For region I, we have by definition $0\leq E_{\bm{p}}\leq E_F$. For region II, we have by definition $E_F\leq E_{\bm{p}}\leq E_F^\lambda$. Also, the final state has energy of $E_{\bm{p}}+\weff$. We now apply these conditions to each transition.
\par \textbf{I$\to$ III:} since $E_{\bm{p}}+\weff\in$ III, $E_F\leq E_{\bm{p}}+\weff\leq E_F^\lambda\Rightarrow E_F-\weff\leq E_{\bm{p}}\leq E_F^\lambda-\weff$. Together with the conditions $\Delta \leq E_{\bm{p}}$, $0\leq E_{\bm{p}}\leq E_F$ we have 
\begin{equation}
\mbox{I$\to$ III:}\quad E_{\mbox{\scriptsize low}}=\mbox{max}\left(\Delta, E_F -\weff\right), \quad E_{\mbox{\scriptsize high}}= \mbox{min}\left(E_F, E_F^\lambda -\weff\right).
\end{equation}
\par \textbf{I$\to$ IV:} since $E_{\bm{p}}+\weff\in$ V, $E_F^\lambda\leq E_{\bm{p}}+\weff\Rightarrow E_F^\lambda -\weff\leq E_{\bm{p}}$. Together with the conditions $\Delta \leq E_{\bm{p}}$, $0\leq E_{\bm{p}}\leq E_F$, we have \begin{equation}
\mbox{I$\to$ IV:}\quad E_{\mbox{\scriptsize low}}=\mbox{max}\left(\Delta, E_F^\lambda -\weff\right), \quad E_{\mbox{\scriptsize high}}= E_F.
\end{equation}
\par \textbf{I$\to$ V:} since $E_{\bm{p}}+\weff\in$ V, $0\leq E_{\bm{p}}+\weff\leq E_F\Rightarrow -\weff\leq E_{\bm{p}}, E_{\bm{p}}\leq E_F-\weff$. Together with the conditions $\Delta \leq E_{\bm{p}}$, $0\leq E_{\bm{p}}\leq E_F$, we have 
\begin{equation}
\mbox{I$\to$ V:}\quad E_{\mbox{\scriptsize low}}=\mbox{max}\left(\Delta,  -\weff\right), \quad E_{\mbox{\scriptsize high}}= \mbox{min}\left(E_F,  E_F-\weff\right).
\end{equation}
\par \textbf{II$\to$ III:} since $E_{\bm{p}}+\weff\in$ III, $E_F\leq E_{\bm{p}}+\weff\leq E_F^\lambda\Rightarrow E_F-\weff\leq E_{\bm{p}}\leq E_F^\lambda-\weff$. Together with the conditions $\Delta \leq E_{\bm{p}}$, $E_F\leq E_{\bm{p}}\leq E_F^\lambda$, we have 
\begin{equation}
\mbox{II$\to$ III:}\quad E_{\mbox{\scriptsize low}}=\mbox{max}\left(\Delta, E_F,E_F -\weff\right), \quad E_{\mbox{\scriptsize high}}= \mbox{min}\left(E_F^\lambda, E_F^\lambda -\weff\right).
\end{equation}
\par \textbf{II$\to$ IV:} since $E_{\bm{p}}+\weff\in$ IV, $E_F^\lambda\leq E_{\bm{p}}+\weff\Rightarrow E_F^\lambda -\weff\leq E_{\bm{p}}$. Together with the conditions $\Delta\leq E_{\bm{p}}$, $E_F\leq E_{\bm{p}}\leq E_F^\lambda$, we have 
\begin{equation}
\mbox{II$\to$ IV:}\quad E_{\mbox{\scriptsize low}}=\mbox{max}\left(\Delta, E_F,E_F^\lambda -\weff\right), \quad E_{\mbox{\scriptsize high}}= E_F^\lambda.
\end{equation}
\par \textbf{II$\to$ V:} since $E_{\bm{p}}+\weff\in$ V, $0\leq E_{\bm{p}}+\weff\leq E_F\Rightarrow -\weff\leq E_{\bm{p}}, E_{\bm{p}}\leq E_F-\weff$. Together with the conditions $\Delta \leq E_{\bm{p}}$, $E_F\leq E_{\bm{p}}\leq E_F^\lambda$, we have 
\begin{equation}
\mbox{II$\to$ V:}\quad E_{\mbox{\scriptsize low}}=\mbox{max}\left(\Delta, E_F, -\weff\right), \quad E_{\mbox{\scriptsize high}}= \mbox{min}\left(E_F^\lambda,  E_F-\weff\right).
\end{equation}

\subsection{Calculation of the cross section}
For each possible transition we use the limits of integration from Sec. \ref{sec:limits} for the $a$ integrals from Eq. (\ref{eq:ai_final}) in Appendix \ref{app:ab_functions}. Using Eq. (\ref{Wi}), we combine those with the components of the nucleon tensor $H_i$ to obtain the final expression for $W_i$. To calculate the cross section, we add all the possible transitions, namely,  
\begin{equation}\label{eq:xs_total}
d\sigma=d\sigma_{I\to III}+d\sigma_{I\to IV}+d\sigma_{I\to V}+d\sigma_{II\to III}+d\sigma_{II\to IV}+d\sigma_{II\to V}.
\end{equation}
The limits of integration ensure that transitions that are not allowed kinematically do not contribute.

\section{RESULTS}\label{sec:results}
Having derived analytical expressions for the cross section, we now compare them to data.  We compare the predictions to electron-carbon data from Refs. \cite{Benhar:2006er,CarbonData, Barreau:1983} and flux-averaged neutrino scattering data from the MiniBooNE experiment \cite{MiniBooNE:2010bsu}. In the following, we focus on the differences between the RFG and CFG models and different form factor parametrizations. 

\subsection{Electron scattering}\label{sec:electron}
We compare the predictions of the RFG and CFG models to electron-carbon scattering data. Since for both models the proton and neutron momentum distributions are independent, we separately obtain the lepton-proton and lepton-neutron scattering cross sections and add them together. Recall that the distributions are normalized to $A/2$, where for carbon $A=12$. 

The difference between the lepton-proton and lepton-neutron scattering cross sections arises from the different electromagnetic form factors of each nucleon.  We compare two different electromagnetic form factor parametrizations: the commonly used Bradford-Bodek-Budd-Arrington (BBBA) parametrization \cite{Bradford:2006yz} and the $z$-expansion-based parametrization from Ref. \cite{Borah:2020gte}, referred to as Borah Hill Lee Tomolak (BHLT) in the following. BHLT is our default parametrization. 
The error bars in the theoretical predictions of the cross section come solely from the uncertainty in form factor models. Here and in Sec. \ref{sec:neutrino}, we use published covariance matrices when available or treat model parameters to be uncorrelated when a covariance matrix is unavailable. We then use standard error propagation with uncorrelated errors added in quadrature.
Details about these parametrizations appear in Appendix \ref{app:form_factors}.

The electron-carbon scattering data is taken from the compilation in Refs. \cite{Benhar:2006er,CarbonData, Barreau:1983}. There are 66 kinematical points corresponding to different values of the initial electron energy and final electron scattering angle. We present a few of them here and include more in Appendix \ref{app:plots}.

\begin{figure}
\begin{center}
\includegraphics[scale=0.6]{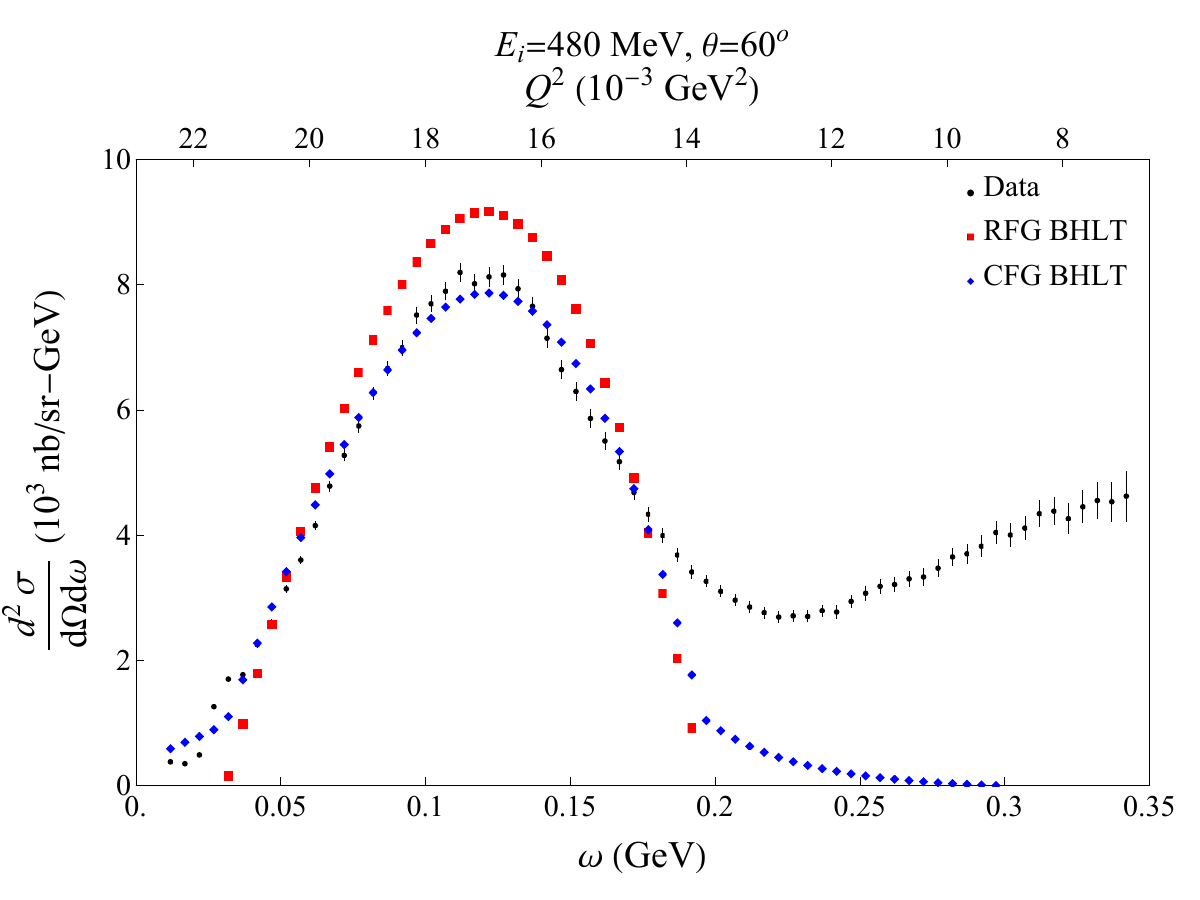}
\caption{\label{Fig:480_MeV_60z} Comparison of RFG (red squares) and CFG (blue diamonds) models to carbon data. The incident electron energy is 480 MeV and the scattering angle is $60^\circ$. The error bars from the form factor uncertainties are smaller than the markers' size of the $\mathcal{O}(10^1 )$ nb/sr-GeV.}
\end{center}
\end{figure}

Consider first a comparison of RFG and CFG models with the BHLT parametrization to carbon data with incident electron energy of 480 MeV and a scattering angle of $60^\circ$; see Fig. \ref{Fig:480_MeV_60z}. In comparing RFG and CFG, we see that the $\omega$ values of the CFG data points extend beyond the RFG data points. These reflect the phase space limits for each model. For the RFG model, the limits for $E_{\bm{p}}$ are $ E_{\mbox{\scriptsize low}}=\mbox{max}\left(\Delta, E_F -\weff\right)$ and  $E_{\mbox{\scriptsize high}}=E_F$. For the considered kinematics, these translate to $0.031\,\mbox{GeV}\leq\omega\leq 0.196\,$GeV.  

\begin{figure}
\begin{center}
\includegraphics[scale=0.6]{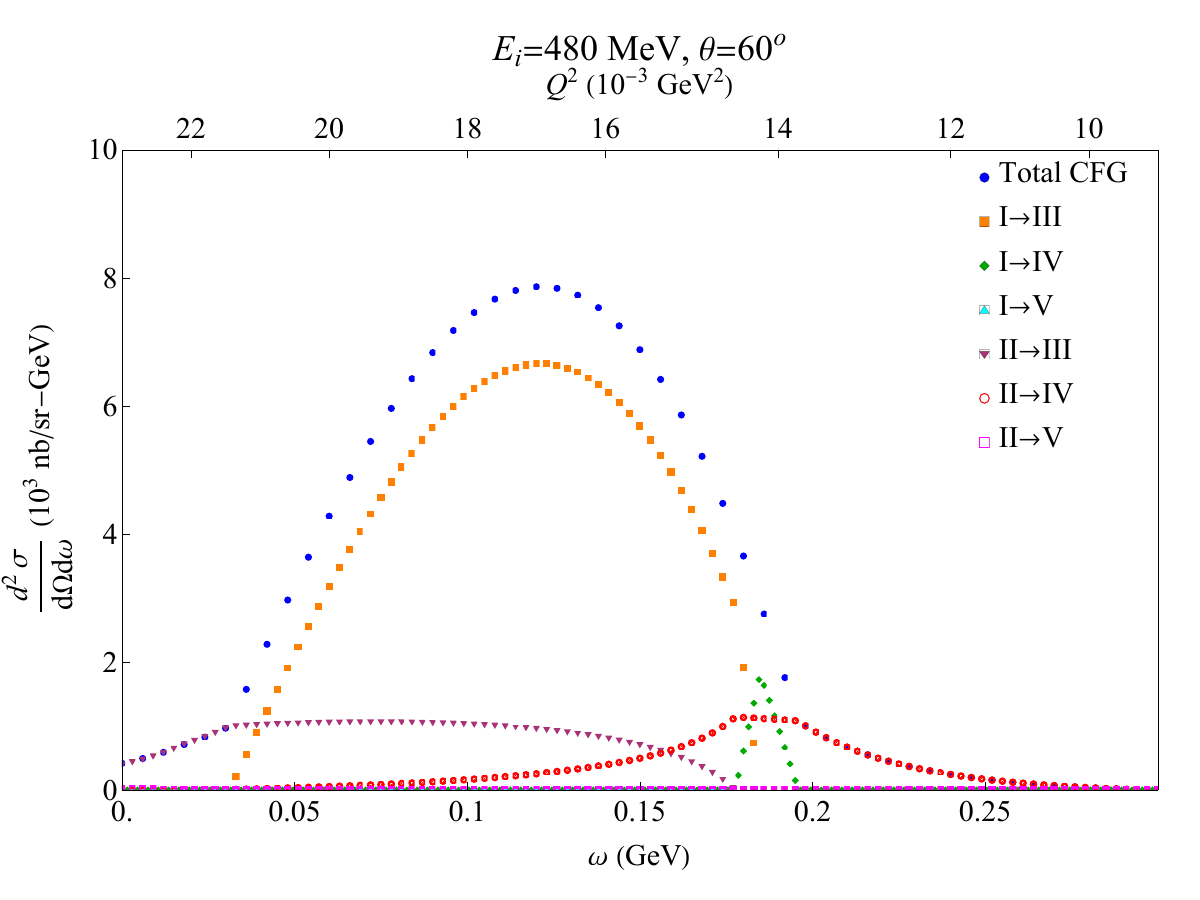}
\caption{\label{Fig:480_MeV_60z_by_Regions} Contributions to the CFG model cross section by transitions. The incident electron energy is 480 MeV and the scattering angle is $60^\circ$.}
\end{center}
\end{figure}

To better understand the CFG case, we show the cross section for this kinematics for each of the six possible transitions in Fig. \ref{Fig:480_MeV_60z_by_Regions}.  The cross section is dominated by transitions from region I to region III. The shape of the cross section for such transitions is analogous to that of the RFG model. The ``tail"  for larger values of $\omega$ is generated by transitions from region I and II to region IV. Only the transition from region II to region IV has values of $\omega$ larger than the RFG case, where $\omega$ can be as large as 0.298 GeV for this kinematic. The tail  for smaller values of $\omega$ is mostly generated by transitions from region II to region III. The transition from region II to V gives a very small contribution, a few percent of the total cross section in the small-$\omega$ region. The transition from region I to V does not contribute for this kinematic. Notice also that different transitions can have the same value of $\omega$, since they will originate from different values of $E_{\bm{p}}$. 

\begin{figure}
\begin{center}
\includegraphics[scale=0.6]{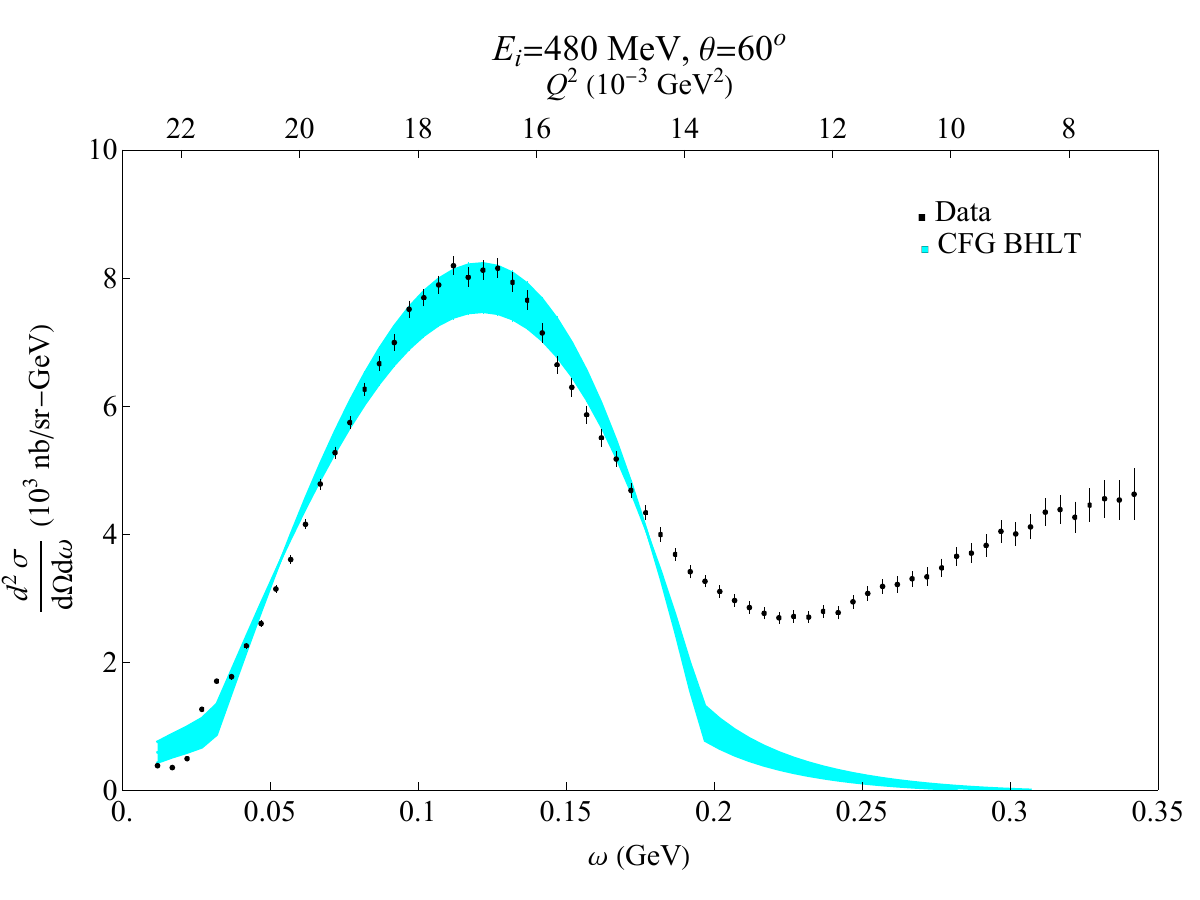}
\caption{\label{Fig:480_MeV_60z_parmtr_varn} Comparison between the carbon data and CFG nuclear model using the BHLT parametrization for values of $\lambda$ and $c_0$ in the range [2.5, 3.0] and [3.21, 5.11], respectively.}
\end{center}
\end{figure}

In plotting the CFG predictions, we use the central values of the model parameters, $\lambda$ and $c_0$. Varying of the model parameters results in broadening of the scattering cross section prediction. We illustrate by the band in Fig. \ref{Fig:480_MeV_60z_parmtr_varn}  for the case of an incoming electron energy of 480 MeV and scattering angle of $60^\circ$. We again use the BHLT form factor parametrization. Comparing to Fig. \ref{Fig:480_MeV_60z}, the variation of the CFG parameters is smaller than the difference between the RFG and CFG models. 

\begin{figure}
\begin{center}
\includegraphics[scale=0.6]{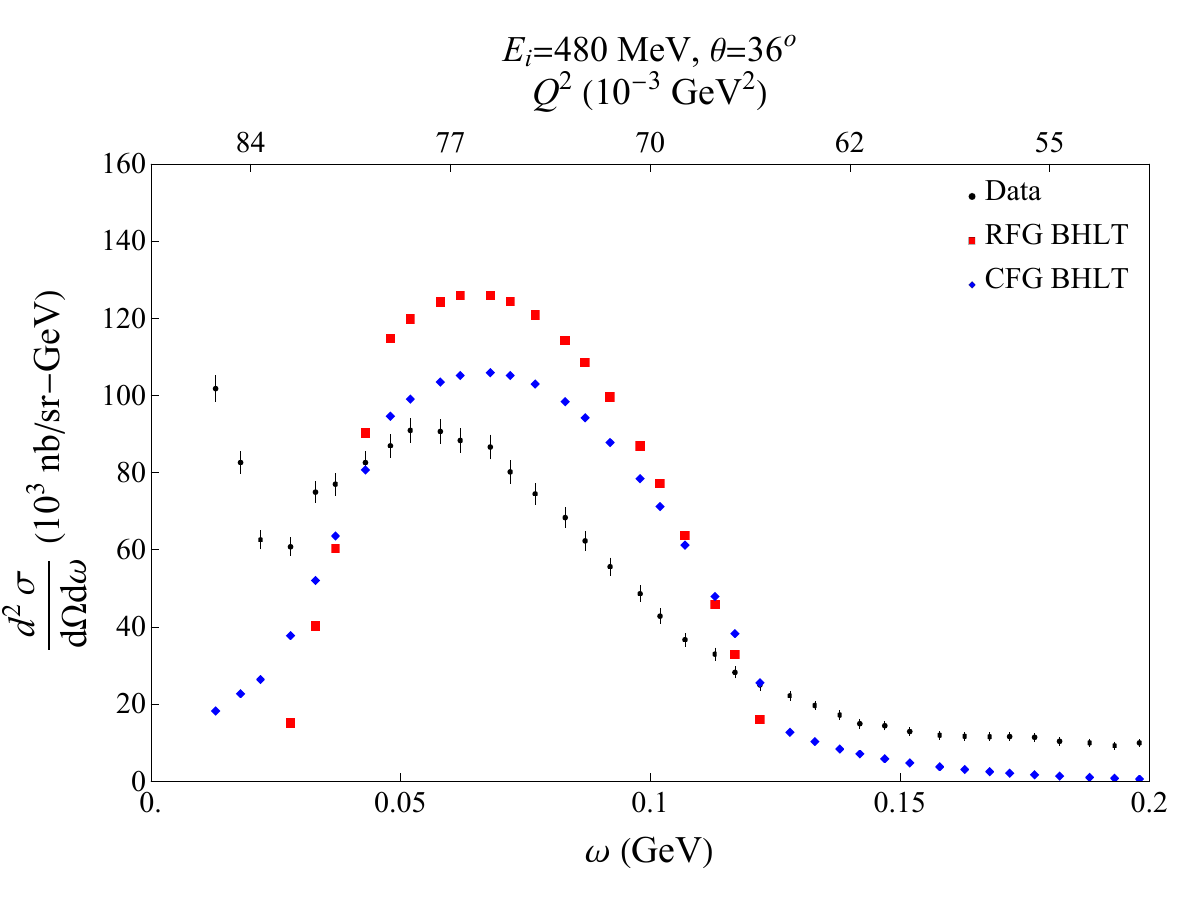}
\caption{\label{Fig:480_MeV_36z} Comparison of RFG (red squares) and CFG (blue diamonds) models to carbon data. The incident electron energy is 480 MeV and the scattering angle is $36^\circ$. The error bars for the RFG and CFG are of the $\mathcal{O}(10^2)$ nb/sr-GeV, which remain invisible due to the size of the markers.}
\end{center}
\end{figure}

 For carbon data with incident electron energy of 480 MeV and a scattering angle of $60^\circ$, the CFG model fits the data better than the RFG model. The same is true if the change the angle the scattering angle to $36^\circ$, but the overall fit is worse; see Fig. \ref{Fig:480_MeV_36z}.

 \begin{figure}[H]
\begin{center}
\includegraphics[scale=0.35]{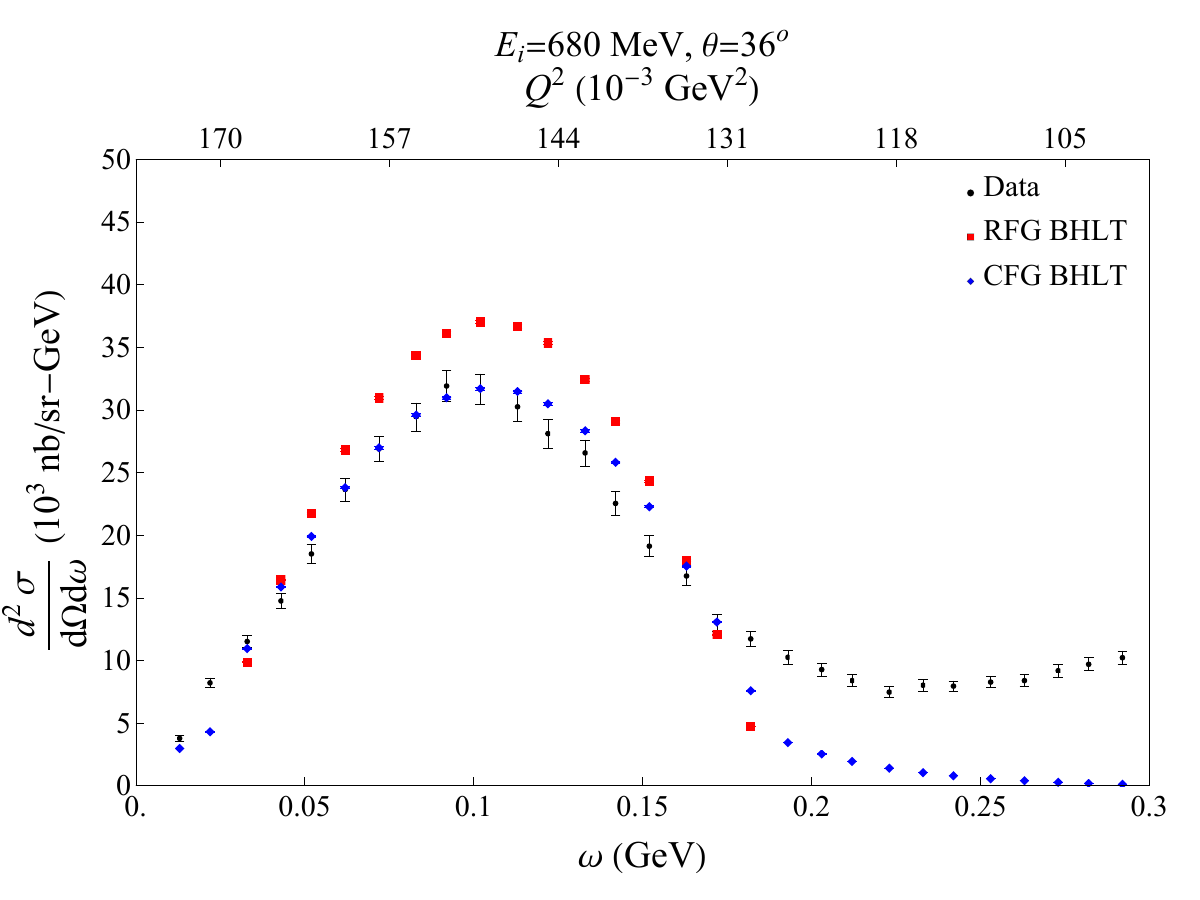}
\hspace{1cm}
\includegraphics[scale=0.35]{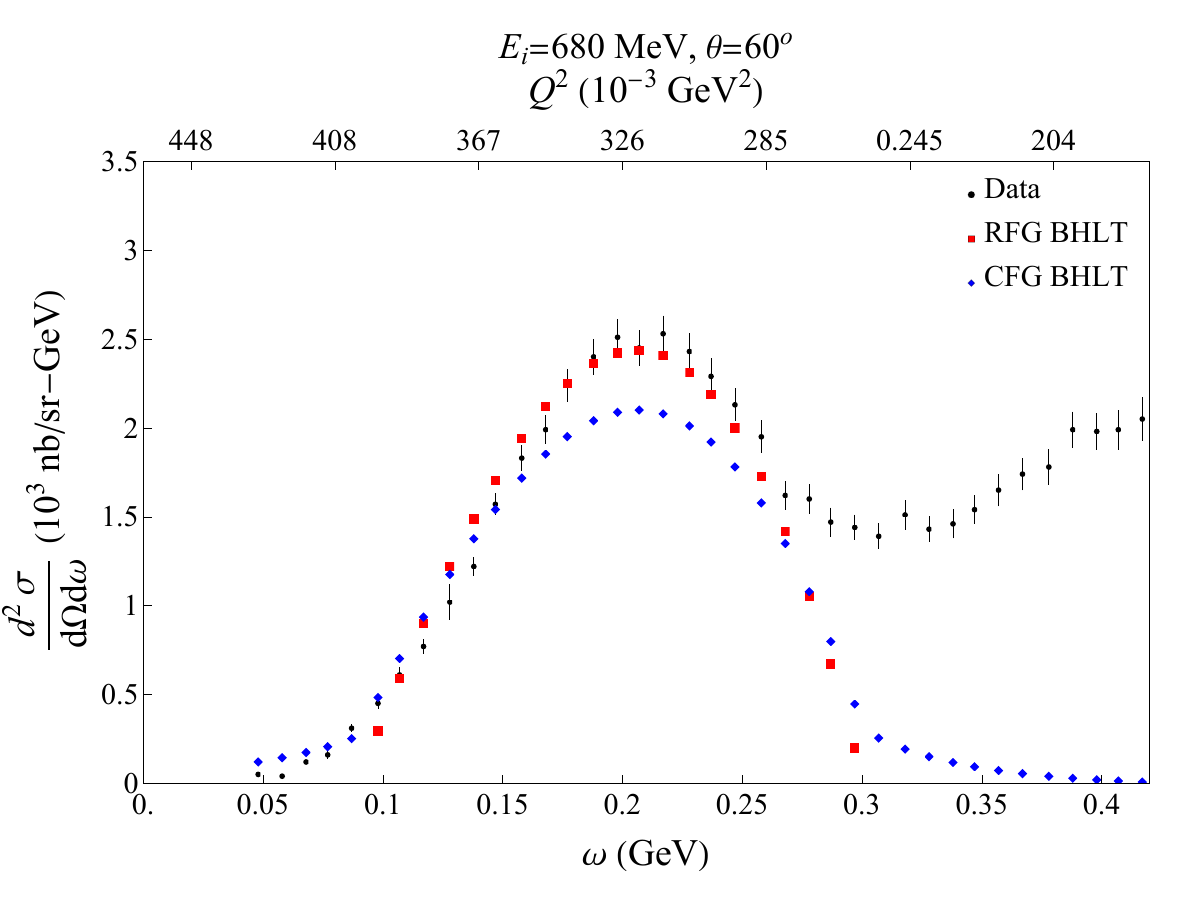}
\caption{\label{Fig:680_MeV} Comparison of RFG (red squares) and CFG (blue diamonds) models to carbon data. The incident electron energy is 680 MeV.  Left: scattering angle of  $36^\circ$. Right: scattering angle of  $60^\circ$. The error bars for the RFG and CFG are of the $\mathcal{O}(10^2)$ nb/sr-GeV for $36^\circ$ and $\mathcal{O}(10^1)$ nb/sr-GeV for $60^\circ$.}
\end{center}
\end{figure}

In Fig. \ref{Fig:680_MeV}, we compare the RFG and CFG models to carbon data with incident electron energy of 680 MeV and a scattering angle of $36^\circ$ (left) and $60^\circ$ (right). For this energy the CFG model fits the data better for $36^\circ$, while the RFG model fits the data better for $60^\circ$.

\begin{figure}[H]
\begin{center}
\includegraphics[scale=0.35]{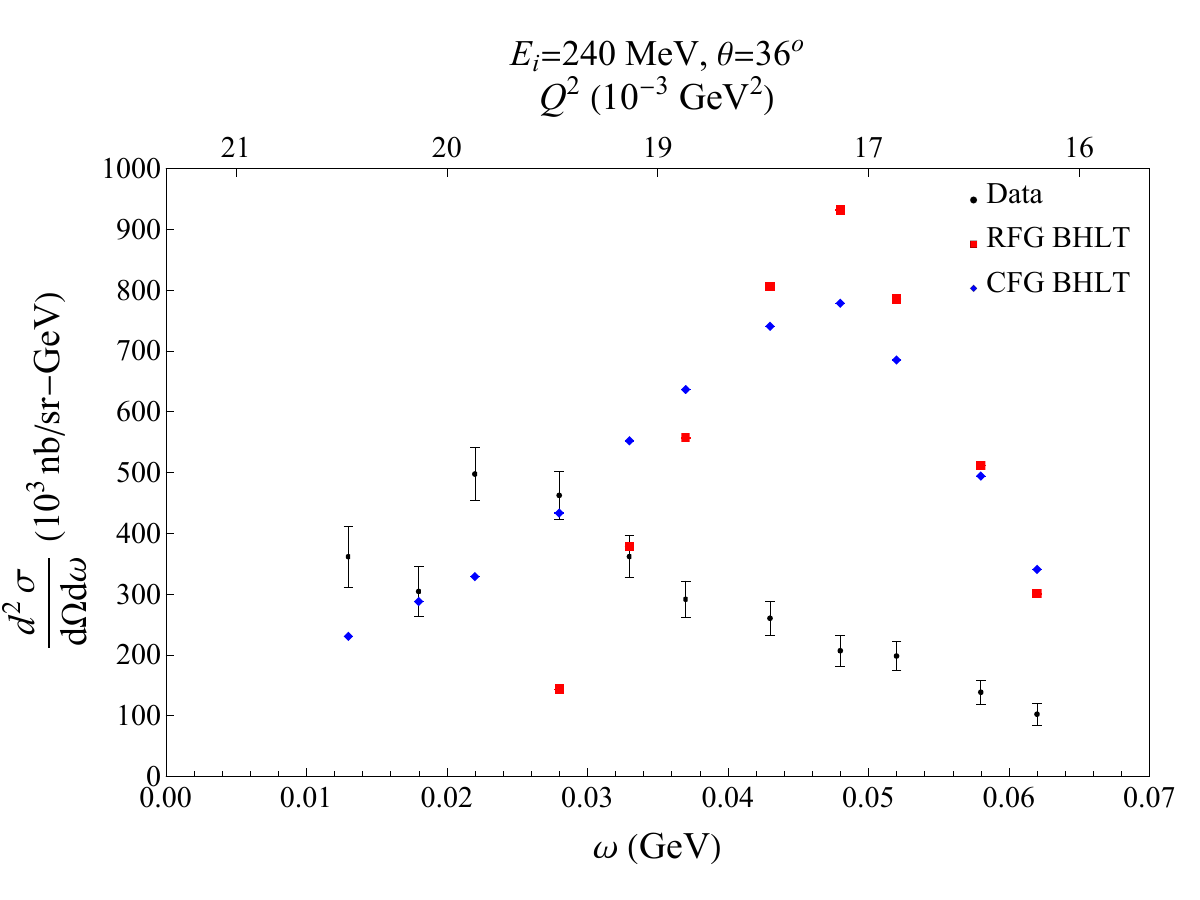}
\hspace{1cm}
\includegraphics[scale=0.35]{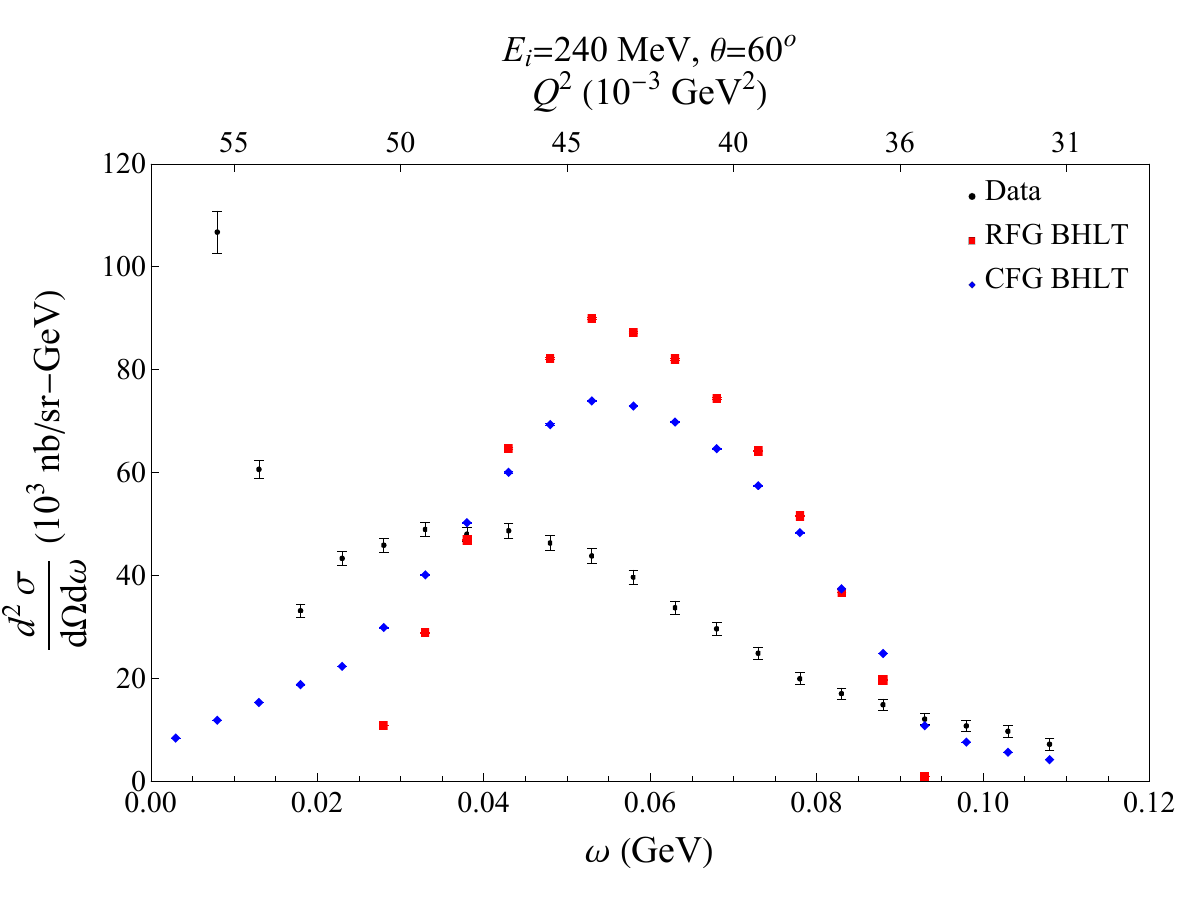}
\caption{\label{Fig:240_MeV} Comparison of RFG (red squares) and CFG (blue diamonds) models to carbon data. The incident electron energy is 240 MeV.  Left: scattering angle of  $36^\circ$. Right: scattering angle of  $60^\circ$.The error bars for the RFG and CFG are of the $\mathcal{O}(10^2)$ nb/sr-GeV with the error being larger for $36^\circ$ compared to $60^\circ$.}
\end{center}
\end{figure}

Finally, we compare the RFG and CFG models to carbon data with incident electron energy of 240 MeV and a scattering angle of $36^\circ$ (left) and $60^\circ$ (right); see Fig. \ref{Fig:240_MeV}.  Neither of the models fits the data very well. In Ref. \cite{Ankowski:2014yfa}, it was shown that a better agreement to the data is obtained by convoluting the cross section with a folding function that describes the effects of Final State Interactions (FSIs) between the outgoing nucleon and remaining nucleus. It would be interesting to include a similar approach for the CFG model.

\begin{figure}[h!]
\begin{center}
\includegraphics[scale=0.35]{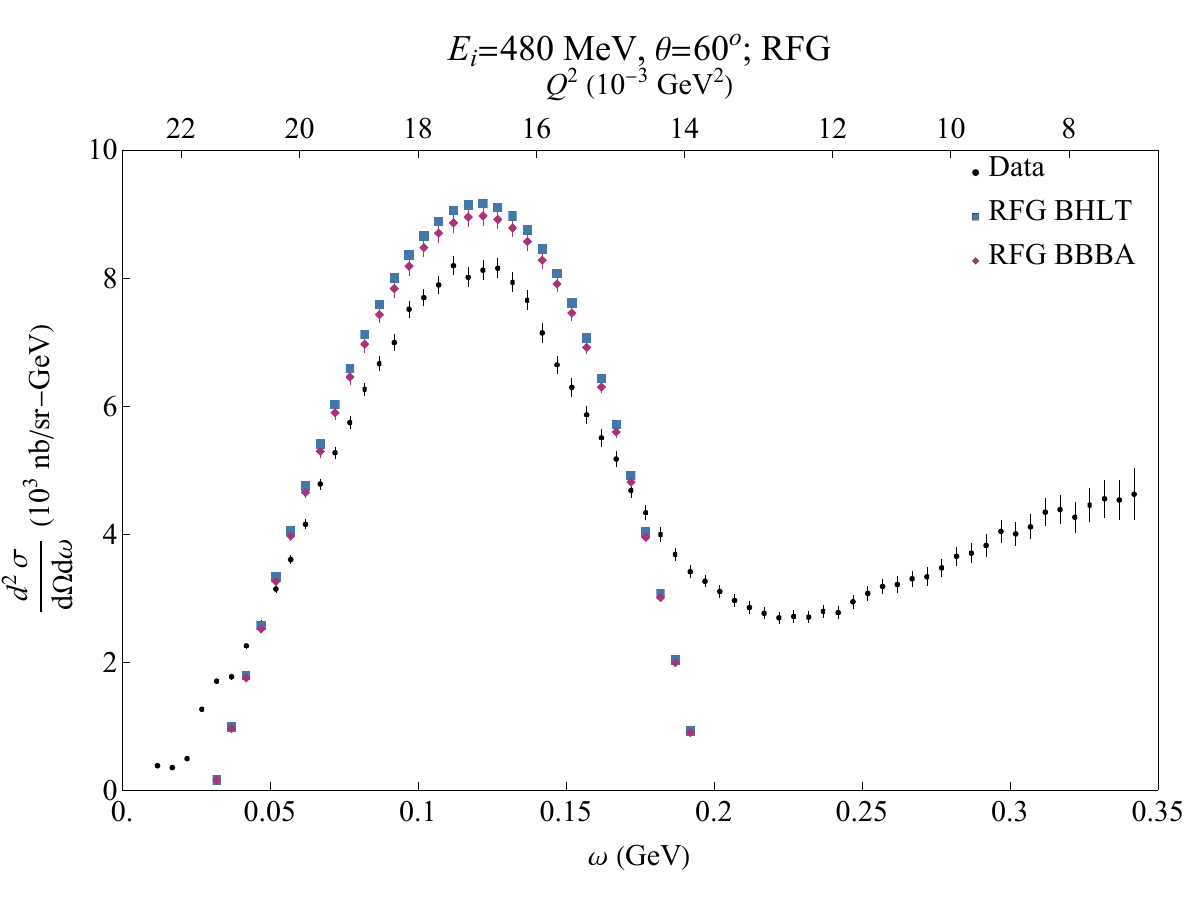}
\hspace{1cm}
\includegraphics[scale=0.35]{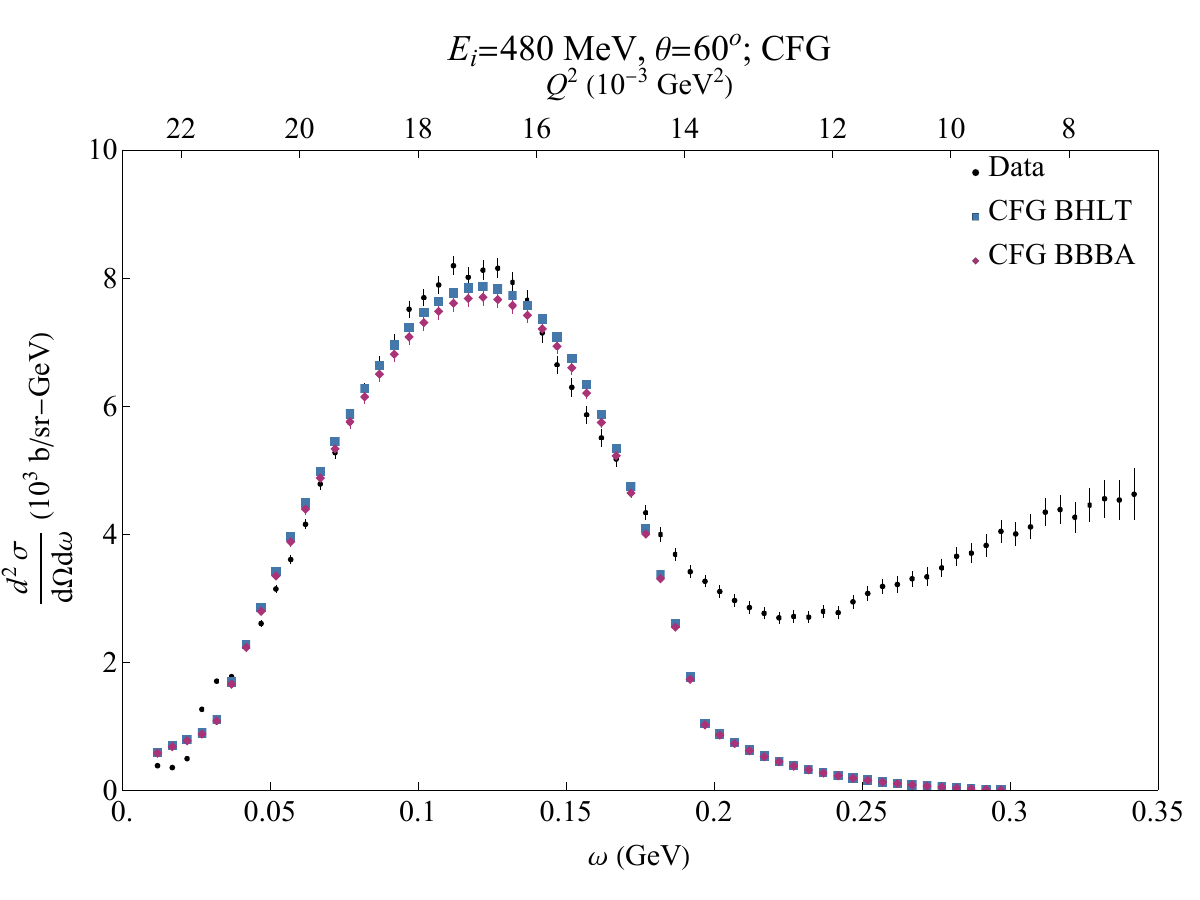}
\caption{\label{Fig:electron_FF_comparison} Comparison of RFG (left) and CFG (right) models to carbon data with the BBBA and $z$-expansion based parametrizations. The incident electron energy is 480 MeV and the scattering angle is $60^\circ$.}
\end{center}
\end{figure}

A different interesting physics question is the separation of nuclear effects, captured by the nuclear model, and nucleon effects, captured by the form factors.  To address this question, we compare the RFG and CFG models to data with two different parametrizations: BBBA \cite{Bradford:2006yz} and BHLT \cite{Borah:2020gte}. In Fig. \ref{Fig:electron_FF_comparison}, we compare each nuclear model using the different parametrizations to carbon data with incident electron energy of 480 MeV and a scattering angle of $60^\circ$. It is clear that the differences between different parametrizations are small compared to the differences between the nuclear models themselves. For example, the maximum value of the differential cross section is at $\omega = 0.122$ GeV. In units of nb/sr-GeV, we have for the RFG model $9159 \pm 37$ (BHLT) and $8968 \pm 146$ (BBBA), while for the CFG model we have $7861 \pm 32$ (BHLT), $7698 \pm 125$ (BBBA).
The results presented here for both RFG and CFG are for a binding energy of 25 MeV. A comparison with a different binding energy value (52.2 MeV) is provided in Appendix~\ref{app:binding energy}.

\subsection{Neutrino scattering}\label{sec:neutrino}
In comparing the CFG model to data, neutrino scattering differs from electron scattering in two important aspects. First, the interaction involves also the axial current apart from the vector current. This requires us to consider five different scalar components of the nuclear tensor: $W_1,\dots,W_5$.  For the CFG model, the extension is clear. We only need to specify two new form factors, $F_A$ and $F_P$, that appear in Eq. (\ref{eq:H_i}). Second, since the incident neutrino energy is not fixed, we need to average over the neutrino flux in order to compare to data.  

Similar to the electromagnetic form factors $F_1$ and $F_2$, we consider several parametrizations for $F_A$. These can be divided to two classes. The first is the historical ``dipole" model 
$F_A^{\rm dipole}(q^2) = F_A(0)/{\left[  1 - {q^2/ (m_A^{\rm dipole})^2}  \right]^2}$, 
where $F_A(0)$ is measured in beta decay \cite{ParticleDataGroup:2022pth}, and $m_A^{\rm dipole}$ is a free parameter.  In the MiniBooNE analysis \cite{MiniBooNE:2010bsu}, $m_A^{\rm dipole}=1.35\pm 0.17$ GeV was obtained, while in the so called ``BBBA07" parametrization \cite{Bodek:2007ym}, $m_A^{\rm dipole}=1.014\pm 0.014$ was obtained. 

The second class are the $z$-expansion based parametrizations. These are more flexible and do not introduce an \textit{a priori} functional form. For the axial form factor the method was pioneered in 2011 \cite{Bhattacharya:2011ah}. Since then, it has been used in the literature to to extract the axial form factor using scattering data and lattice QCD. In the following, we list extractions that give the $z$-expansion coefficients and their uncertainties. In 2016, the axial form factor was extracted from neutrino-deuteron scattering data \cite{Meyer:2016oeg}. In 2023, it was extracted from antineutrino-proton scattering
 by the MINERvA experiment \cite{MINERvA:2023avz}.

 Recently, several lattice QCD $z$-expansion based parametrizations of the axial form factor that give the coefficients and their uncertainties became available. These are by the the Regensburg lattice QCD Group (RQCD)  in 2020 \cite{RQCD:2019jai}, Nuclear Matrix Element (NME) Collaboration in 2021 \cite{Park:2021ypf}, the Mainz group in 2022 \cite{Djukanovic:2022wru}, and in 2023 by the Precision Nucleon Decay Matrix Elements (PNDME) Collaboration \cite{Jang:2023zts} and the Extended Twisted Mass (ETMC) Collaboration \cite{Alexandrou:2023qbg}.

In Appendix \ref{app:form_factors}, we plot the axial form factor from these various parametrizations. Similar plots are available in Ref. \cite{Tomalak:2023pdi}. These plots suggest that Refs. \cite{Meyer:2016oeg} and \cite{Djukanovic:2022wru} represent two extremes of possible parametrizations of $F_A$, with other parametrizations lying in between them. In the following we refer to Ref. \cite{Meyer:2016oeg} as ``MBGH" and Ref. \cite{Djukanovic:2022wru} as ``Mainz22," and use them to illustrate the possible range of axial form factor uncertainty.  

 Effects from $F_P$ are suppressed by $m_\ell ^2/m_T^2$; see Eqs. (\ref{eq:H_i}) and (\ref{eq:xs_neutrino}). Because of that, we use the pion-pole approximation $F_P(q^2) \approx 2 m_N^2 F_A(q^2)/ (m_\pi^2 -q^2)$. There are extractions of $F_P$ from lattice QCD in the references above, and one could potentially use them instead. 

\subsubsection{Neutrino cross section before flux averaging}
Before comparing the CFG model predictions to MiniBooNE data, let us consider the hypothetical case of a fixed neutrino energy. Neutrino scattering experiments typically have a distribution of energies. We choose a neutrino energy of $E_\nu=1$ GeV, around the peak energy of the MiniBooNE neutrino flux. 

\begin{figure}[H]
\begin{center}
\includegraphics[scale=0.35]{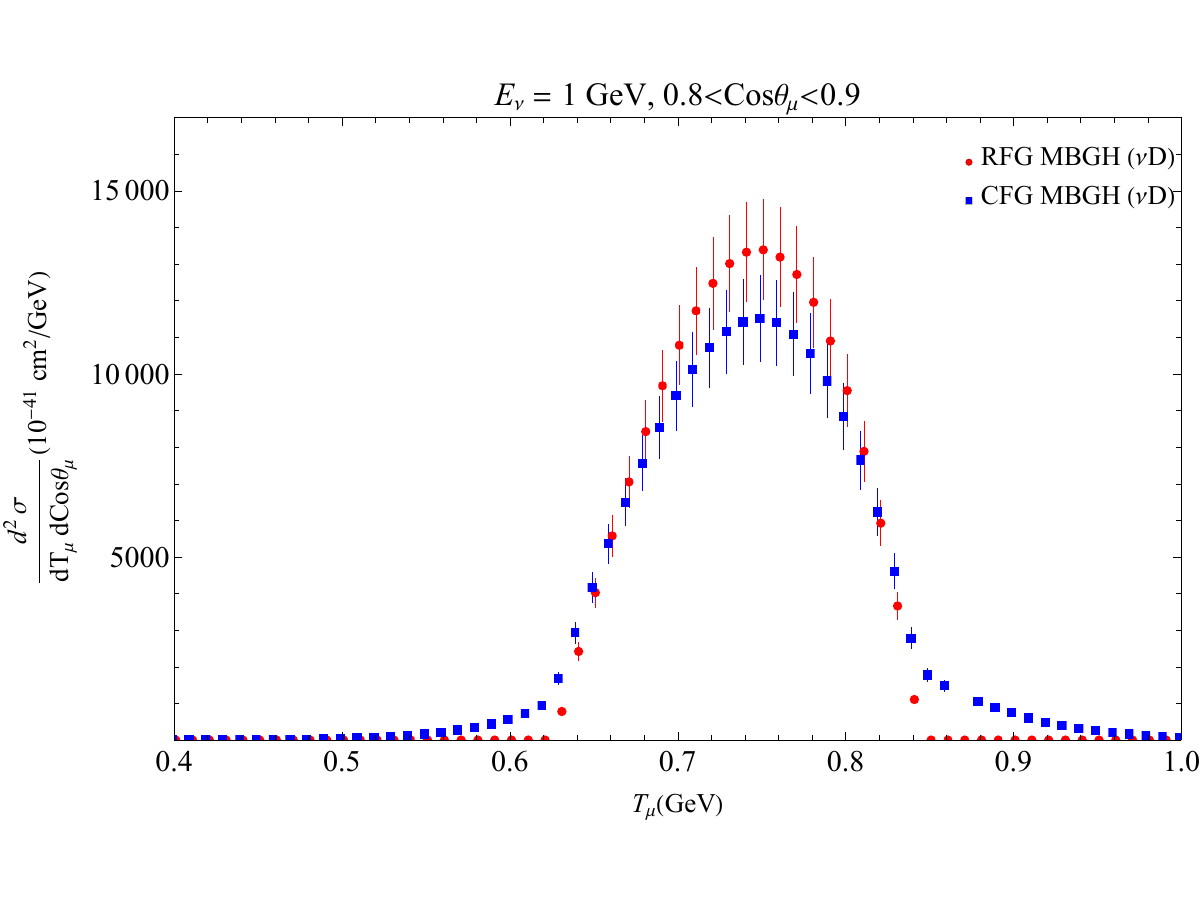}
\hspace{1cm}
\includegraphics[scale=0.35]{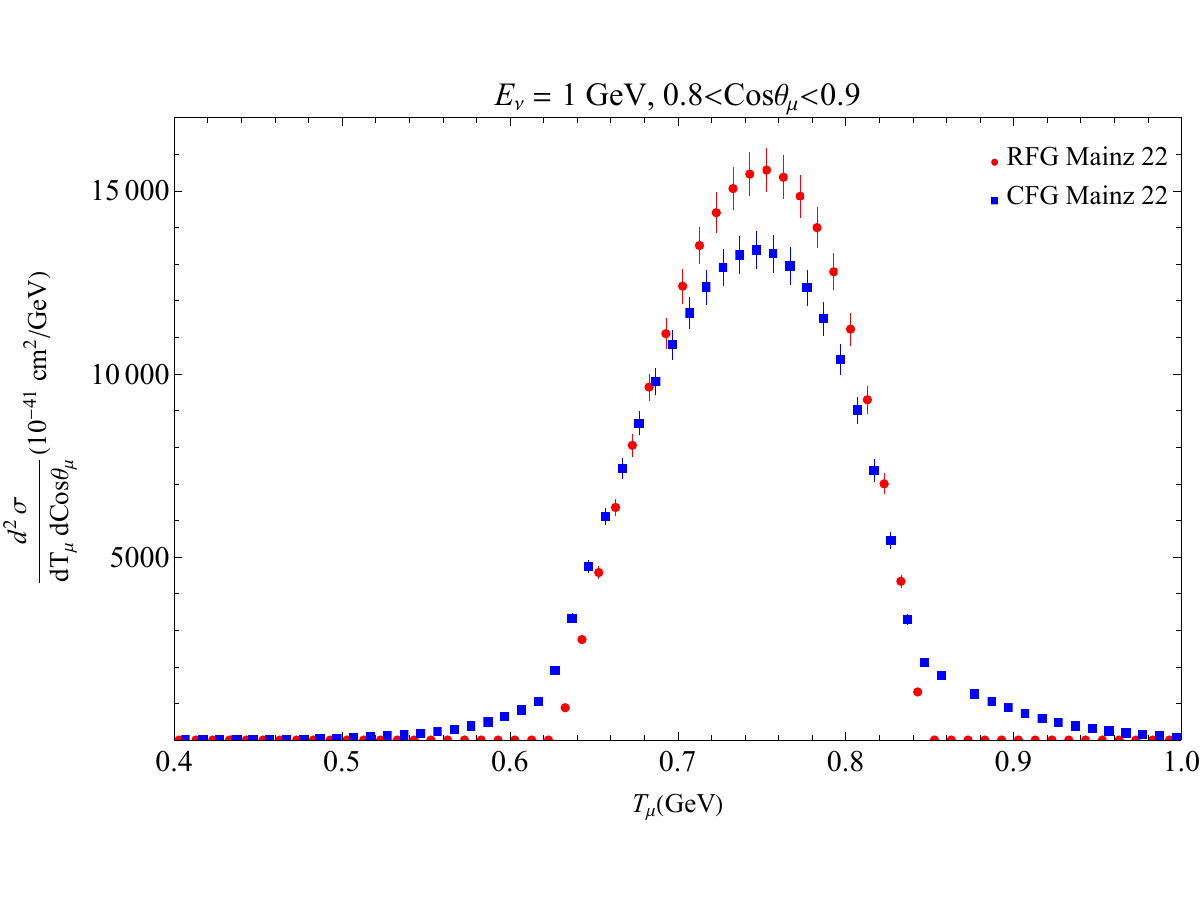}
\caption{\label{Fig:Fixed_Energy0.85} Comparison between RFG (red circles) and CFG (blue squares) model double differential scattering cross section using fixed incoming neutrino energy of $1000$ MeV as a function of the outgoing muon kinetic energy scattering at an angle of $\cos\theta_\mu = 0.85$ Left: MBGH axial form factor. Right: Mainz22 axial form factor.}
\end{center}
\end{figure}

\begin{figure}[H]
\begin{center}
\includegraphics[scale=0.35]{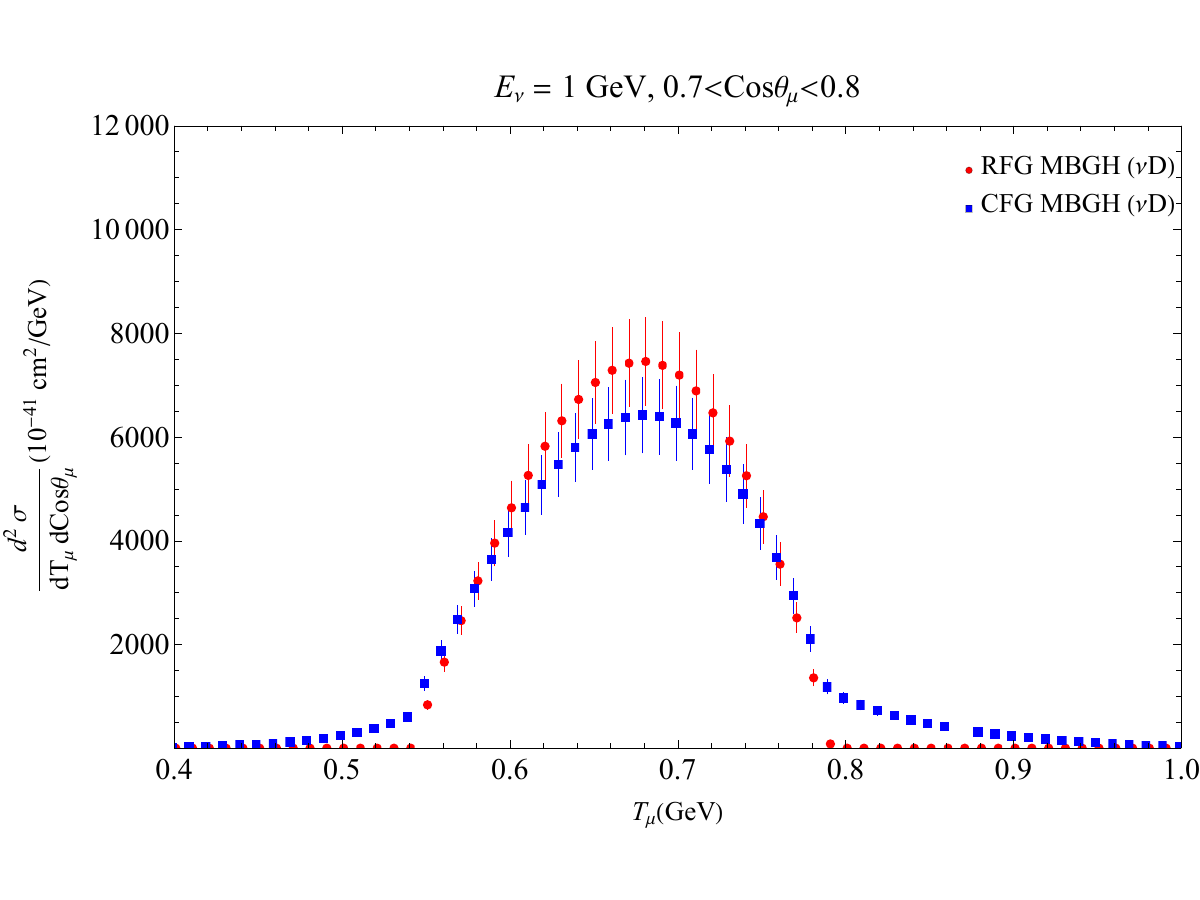}
\hspace{1cm}
\includegraphics[scale=0.35]{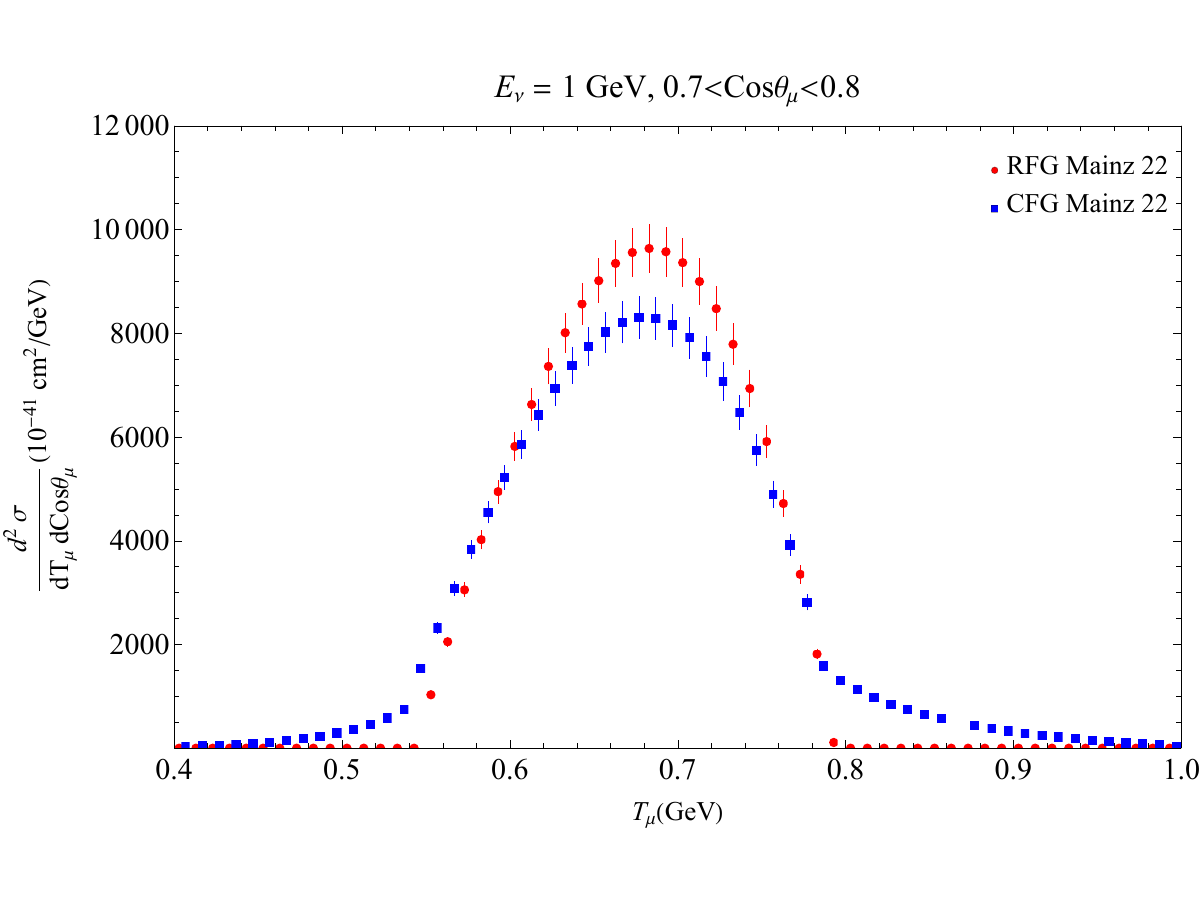}
\caption{\label{Fig:Fixed_Energy0.75} Comparison between RFG (red circles) and CFG (blue squares) model double differential scattering cross section using fixed incoming neutrino energy of $1000$ MeV as a function of the outgoing muon kinetic energy scattering at an angle of $\cos\theta_\mu = 0.75$ Left: MBGH axial form factor. Right: Mainz22 axial form factor.}
\end{center}
\end{figure}

 We compare the RFG and CFG model predictions for the differential cross section of scattering of a 1 GeV neutrino off carbon for $\cos\theta_\mu = 0.85$ (Fig. \ref{Fig:Fixed_Energy0.85}) and $\cos\theta_\mu = 0.75$ (Fig. \ref{Fig:Fixed_Energy0.75}). The predictions are plotted as a function of the muon kinetic energy $T_\mu=E_\mu-m_\mu$. For a fixed neutrino energy, the relation between $\omega$ and $T_\mu$ is $T_\mu=E_\nu-m_\mu-\omega$.  We use BHLT  parametrization for the vector form factors and compare MBGH and Mainz22 for the axial form factor. At the tail region of high or low $T_\mu$, we can clearly distinguish between the RFG and CFG model predictions, independent of the axial form factor parametrization. At the ``peak" region, we can distinguish between RFG and CFG model predictions only for the Mainz22 parametrization. For the MBGH parametrization, the uncertainties overlap between the two nuclear models.

\subsubsection{Neutrino cross section after flux averaging}

We now consider predictions where the neutrino cross section is convoluted with the neutrino flux distribution. In particular, we have \cite{Bhattacharya:2015mpa}

\begin{equation}\label{xscarbonavg}
\frac{d\sigma_{\rm carbon, per\, nucleon, avg.}}{dE_\ell d\cos\theta_\ell}=\int dE_{\nu}\,f(E_{\nu})\,\frac{d\sigma_{\rm carbon, per\, nucleon }}{dE_\ell d\cos\theta_\ell}.
\end{equation}

We compare theoretical predictions to the published MiniBooNE data \cite{MiniBooNE:2010bsu} ``as is" to illustrate the differences between the RFG and CFG models and different form factor parametrizations. For the theory error bars, we follow the same technique as explained in Sec. \ref{sec:electron}. The uncertainty on the MiniBooNE data is obtained by adding in quadrature the shape uncertainty and the 10.7\% normalization uncertainty \cite{MiniBooNE:2010bsu}. See Ref. \cite{Avanzini:2021qlx} for a critical discussion of the MiniBooNE data.

\begin{figure}[H]
\begin{center}
\includegraphics[scale=0.35]{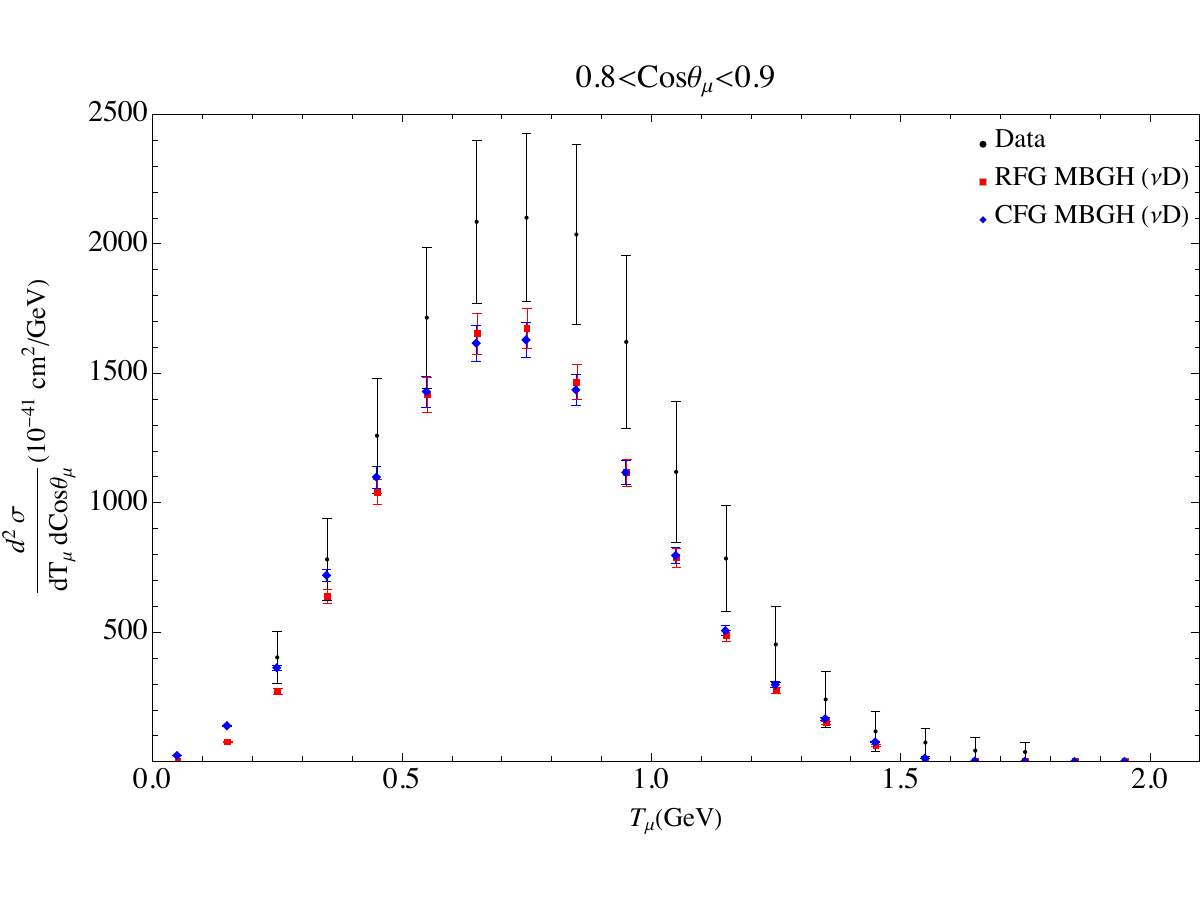}
\hspace{1cm}
\includegraphics[scale=0.35]{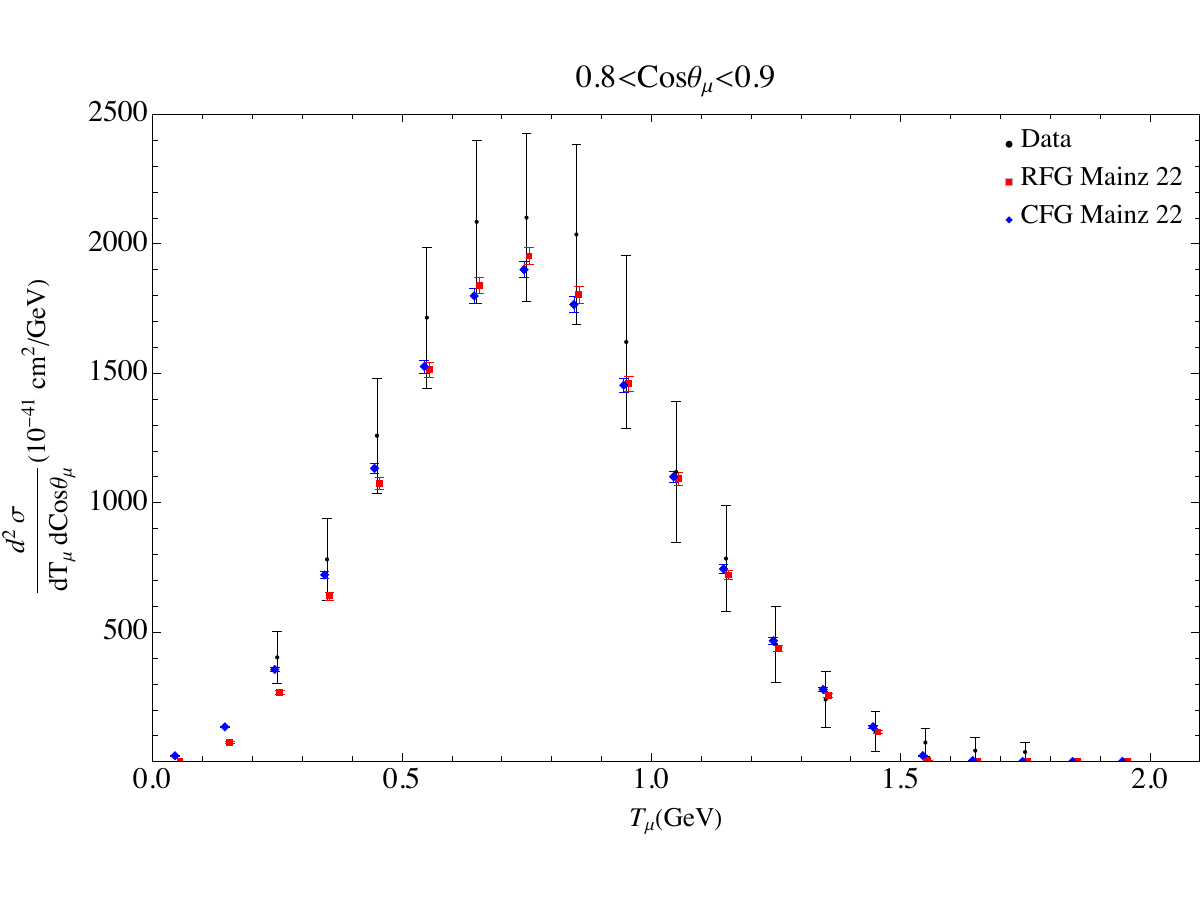}
\caption{\label{Fig:Theta_0.85} Comparison of RFG (red squares) and CFG (blue diamonds) models to flux-averaged neutrino - carbon scattering data as a function of the outgoing muon kinetic energy scattering at an angle of $0.8 < \cos\theta_\mu < 0.9$ using the MBGH axial form factor (left) and Mainz22 for the axial form factor (right). BHLT parametrization is used for the electric and magnetic form factors in both cases.}
\end{center}
\end{figure}

\begin{figure}[H]
\begin{center}
\includegraphics[scale=0.35]{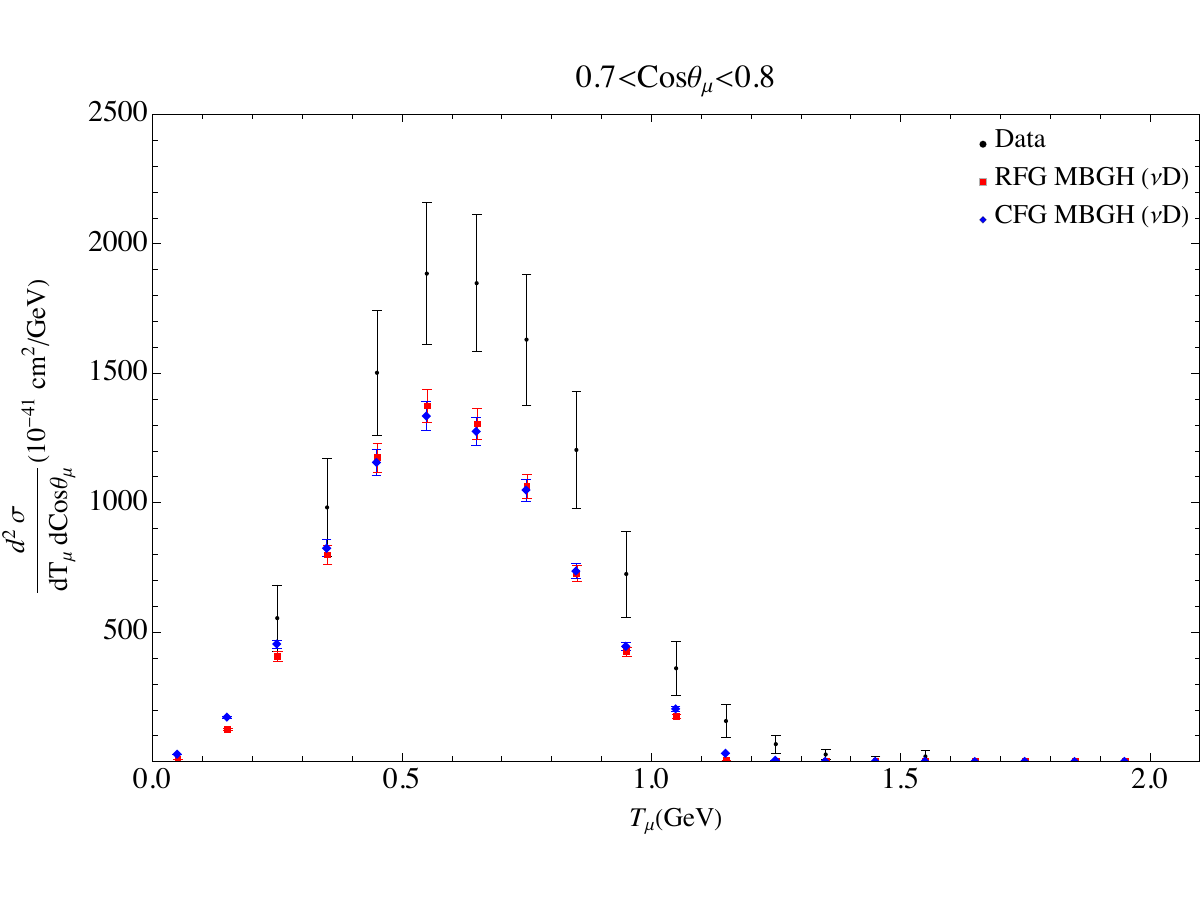}
\hspace{1cm}
\includegraphics[scale=0.35]{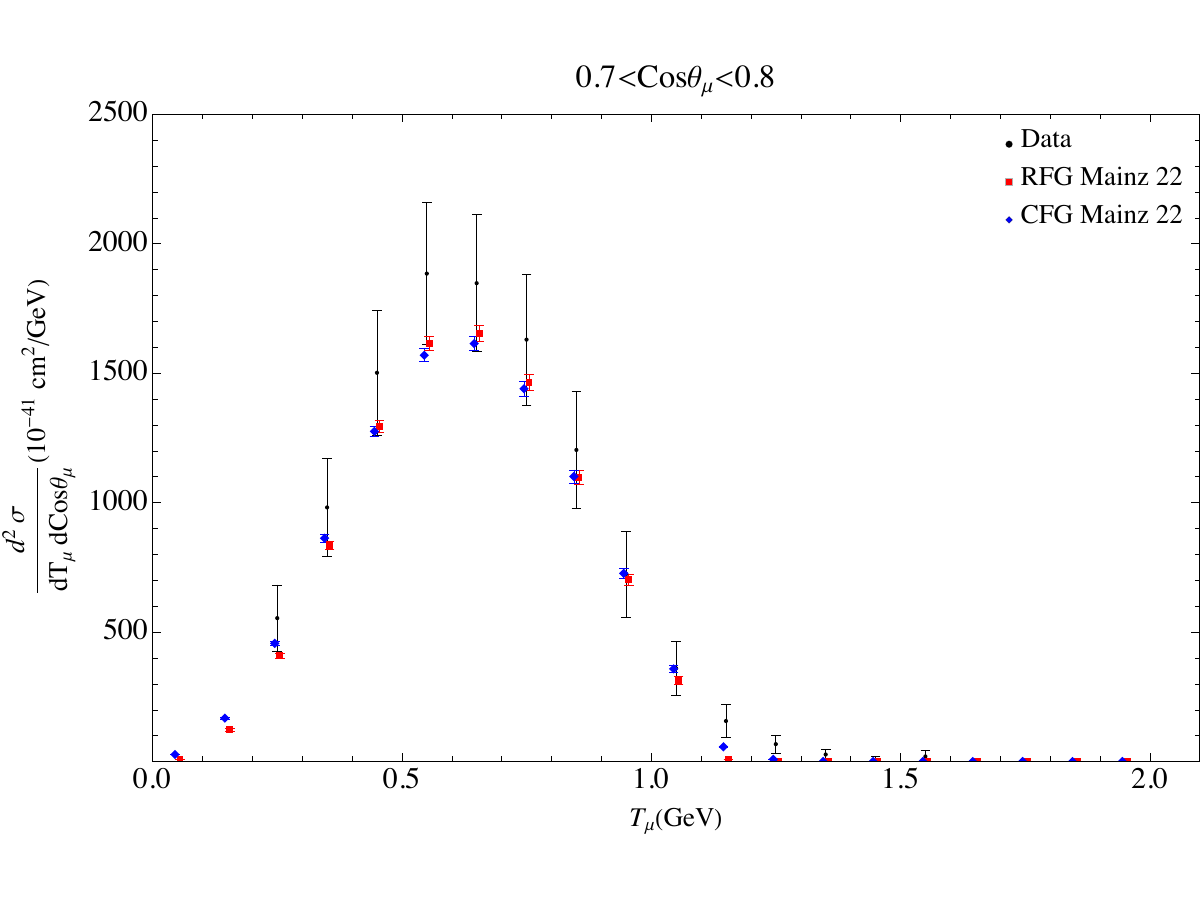}
\caption{\label{Fig:Theta_0.75} Comparison of RFG (red squares) and CFG (blue diamonds) models to flux-averaged neutrino - carbon scattering data as a function of the outgoing muon kinetic energy scattering at an angle of $0.7 < \cos\theta_\mu < 0.8$ using the MBGH axial form factor (left) and Mainz22 for the axial form factor (right). BHLT parametrization is used for the electric and magnetic form factors in both cases.}
\end{center}
\end{figure}

We first consider a \emph{fixed} axial form factor parametrization and \emph{vary} the nuclear models. In Figs. \ref{Fig:Theta_0.85} and \ref{Fig:Theta_0.75}, we show the comparison of the RFG and CFG models to the MiniBooNE data 
for two different bins of $\cos\theta$ and different values of $T_\mu=E_\mu-m_\mu$. 

Unlike the scattering of neutrinos with fixed energy, for the flux-averaged cross section, we cannot distinguish between the RFG and CFG models. The reason is that the averaging over the flux ``adds" several cross sections that peak at different values of $\omega$ ,``smearing" the tails of the cross section. This conclusion is true for either choice of the two $F_A$ parametrizations of MBGH and Mainz22. We also see that the Mainz22 $F_A$ fits the data much better than MBGH. This is not surprising, considering that the Mainz22 $F_A$ largely overlaps with the dipole $F_A$ with $m_A^{\rm dipole}$ extracted by MiniBooNE \cite{MiniBooNE:2010bsu}, assuming the RFG model  without the addition of multinucleon processes \cite{Martini:2009, Martini:2010, Benhar:2010, Barbaro:2011, Amaro:2011, Nieves:2011, Nieves:2011b, Bodek:2011, Meucci:2011}; see the left-hand side of Fig. \ref{Fig:FA_Cmprsn}.

This is a somewhat disappointing result. At least for the MiniBooNE data, we cannot distinguish between the two models.  It would be interesting to see if this phenomenon still persists in less inclusive observable, e.g., semi-inclusive neutrino scattering.

\begin{figure}[H]
\begin{center}
\includegraphics[scale=0.35]{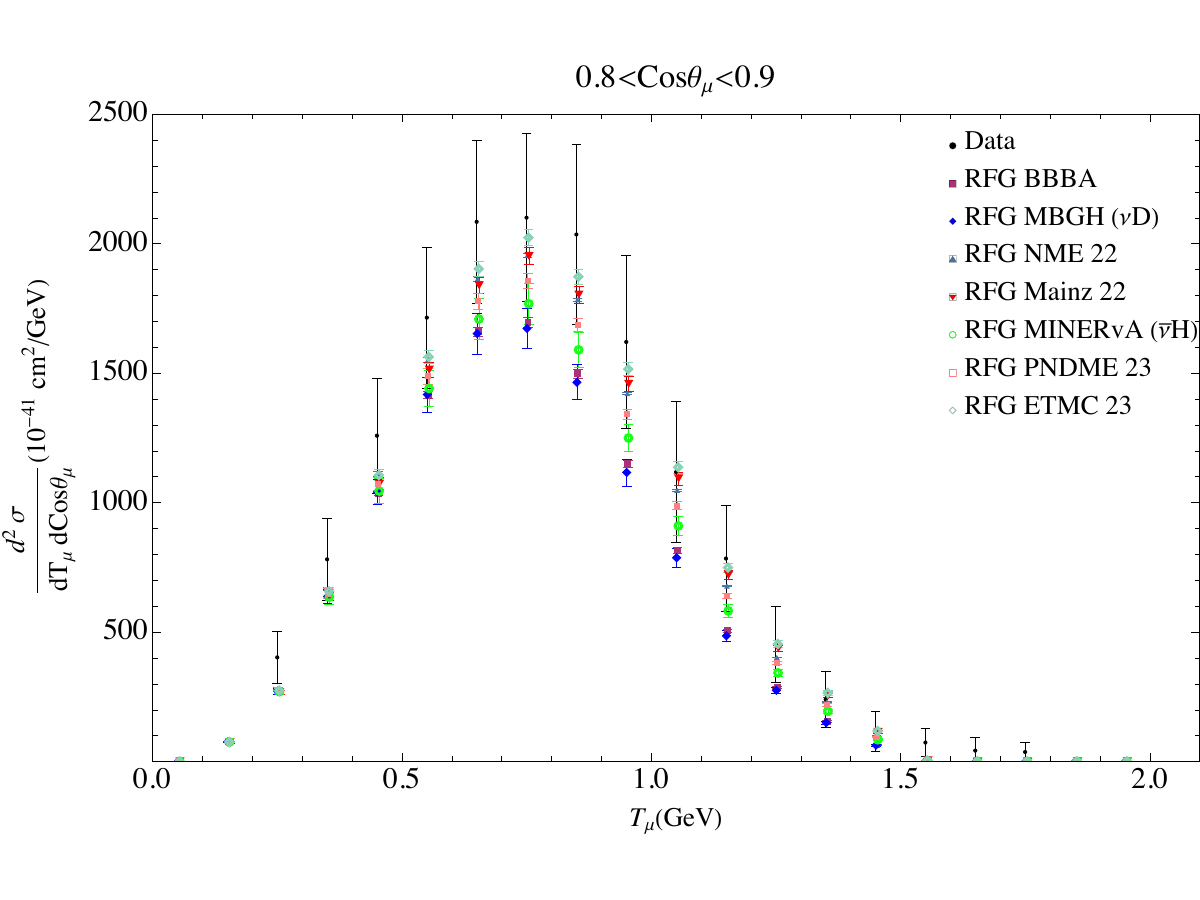}
\hspace{1cm}
\includegraphics[scale=0.35]{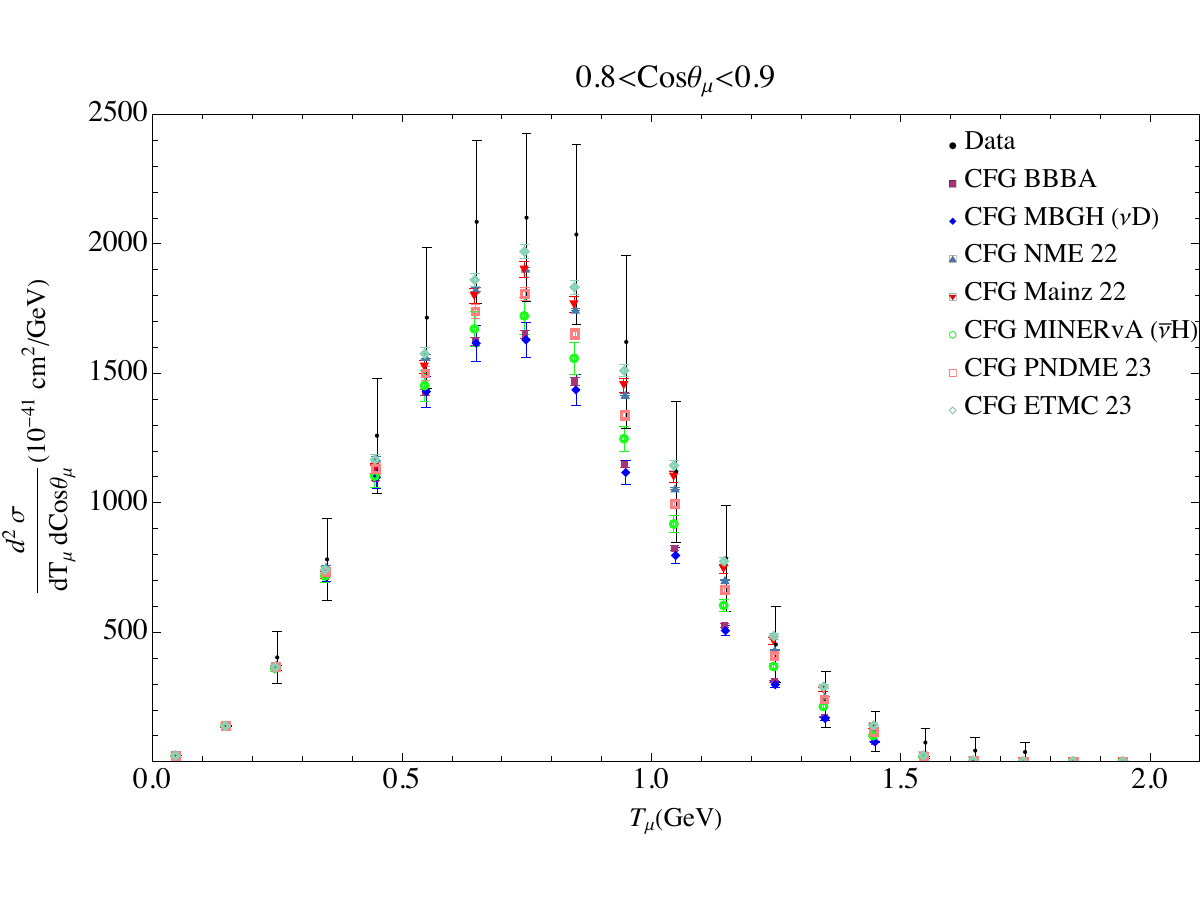}
\caption{\label{Fig:Theta_0.85_RFG_CFG} Comparison of BBBA, MBGH, NME 22, Mainz22, MINERvA ($\bar\nu$H), PNDME 23, and ETMC 23  form factor parametrization to flux-averaged neutrino - carbon scattering data as a function of the outgoing muon kinetic energy scattering at an angle of $0.8 < \cos\theta_\mu < 0.9$ using the RFG model (left) and CFG model (right). BHLT parametrization is used for the electric and magnetic form factors in all the cases except for BBBA.}
\end{center}
\end{figure}

\begin{figure}[H]
\begin{center}
\includegraphics[scale=0.35]{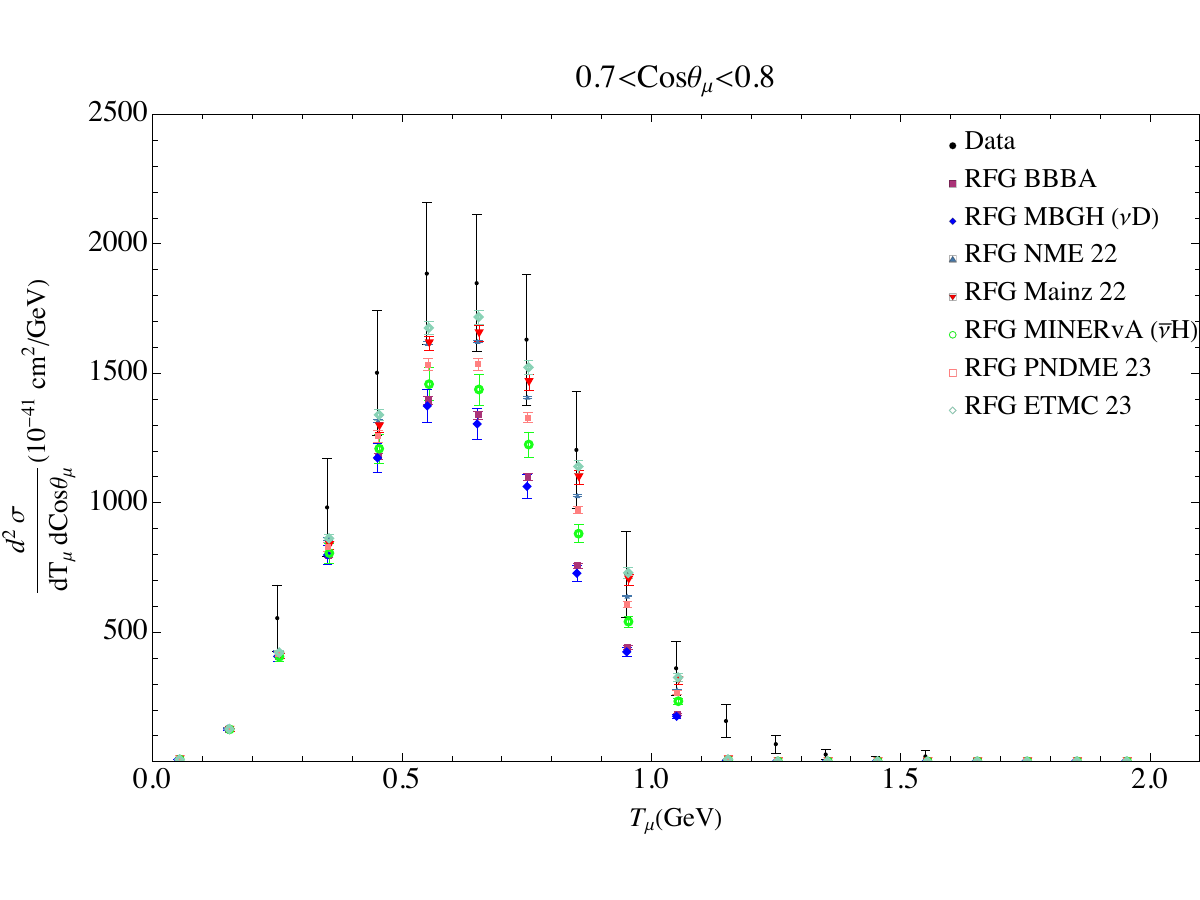}
\hspace{1cm}
\includegraphics[scale=0.35]{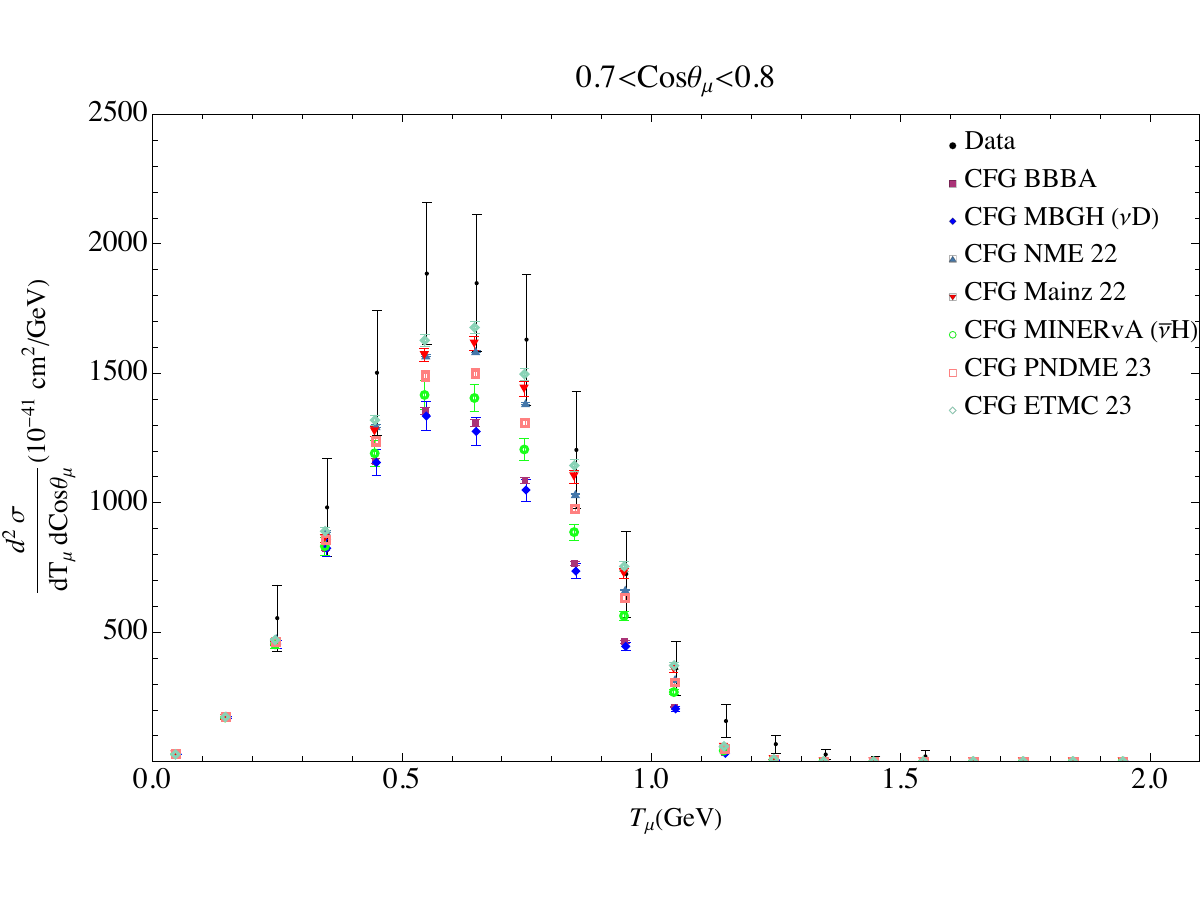}
\caption{\label{Fig:Theta_0.75_RFG_CFG} Comparison of BBBA, MBGH, NME 22, Mainz22, MINERva ($\bar\nu$H), PNDME 23, and ETMC 23  form factor parametrization to flux-averaged neutrino-carbon scattering data as a function of the outgoing muon kinetic energy scattering at an angle of $0.7 < \cos\theta_\mu < 0.8$ using the RFG model (left) and CFG model (right). BHLT parametrization is used for the electric and magnetic form factors all the cases except NME and BBBA.}
\end{center}
\end{figure}

Next we  \emph{vary} the axial form factor parametrization and consider a \emph{fixed} nuclear model. In Figs. \ref{Fig:Theta_0.85_RFG_CFG} and \ref{Fig:Theta_0.75_RFG_CFG},  we compare the effects of a wide range of $F_A$ parametrizations on the neutrino cross section for both RFG and CFG models.  As expected from the right-hand side of Fig. \ref{Fig:FA_Cmprsn},  there is almost a continuous ``spread" from the $F_A$ parametrizations for both nuclear models. We conclude that the axial form factor uncertainty is the dominant one. 

A comparison of axial form factor extractions from lattice QCD and data available before 2022 was presented in Ref. \cite{Meyer:2022mix}. Nuclear effects were not considered in Ref. \cite{Meyer:2022mix}. The authors of Ref.~\cite{Simons:2022ltq} looked at the effect of varying the axial form factor parametrizations: dipole, MBGH, RQCD, and Mainz22 on the flux-averaged charged-current neutrino-nucleus data from the MiniBooNE and T2K Collaborations. The axial form factors were combined with GFMC method and the spectral function formalism.
The flux-averaged cross sections presented here for both RFG and CFG are for a binding energy of 25 MeV. A comparison with a different binding energy value (52.2 MeV) using the MBGH parametrization is provided in Appendix~\ref{app:binding energy}.

\section{Summary and outlook}\label{sec:Summary}

The neutrino research program in the coming decades will require improved precision. A major source of uncertainty is the interaction of neutrinos with nuclei that serve as the target of many such experiments. Broadly speaking, this interaction often depends, e.g., for CCQE, on the combination of nucleon physics, expressed by form factors, and nuclear physics, expressed by a nuclear model. It is important to get a good handle on both. 

In this paper, we presented a fully analytic implementation of the correlated Fermi gas model for CCQE electron-nucleus and neutrino-nucleus scattering. We then used this implementation to compare \emph{separately} form factors and nuclear model effects for both electron-carbon and neutrino-carbon scattering data.

For CCQE in the CFG model, the initial nucleon can be in two possible momentum regions that we label I and II, and the final nucleon can be in three momentum regions that we label III, IV, and V; see Fig. \ref{Fig:np}. The total cross section is given as a sum of the possible six transitions; see Eq. (\ref{eq:xs_total}). The differential cross section for each transition is expressed by the component of the nuclear tensor $W_i$; see Eqs. (\ref{eq:xs_neutrino}) and (\ref{eq:xs_electron}). $W_i$ for each possible transition is expressed as a sum of products of a single-nucleon tensor components and a phase space integrals $a_{1,\dots,7}$; see Eq. (\ref{Wi}). The single-nucleon tensor components are given in Eq. (\ref{eq:H_i}). The phase space integrals for each transition are given in Eq. (\ref{eq:ai_final}).
They can be expressed by a smaller set of master integrals; see Eq. (\ref{eq:bij}). The limits of the integrals are given in Sec. \ref{sec:limits}. Combining all these elements gives an explicit analytical expression for the differential cross section.  

Using these analytical expressions, we compared the CFG model prediction in Sec. \ref{sec:results} to  electron-carbon data and  neutrino scattering data from the MiniBooNE experiment. We focused on the differences between the RFG and CFG models and the effects of different form factor parametrizations.    

In Sec. \ref{sec:electron} we compared the predictions of the RFG and CFG models to electron-carbon scattering data. We have used the BHLT \cite{Borah:2020gte} parametrization of the vector form factors as our default. In all cases, one can clearly distinguish the two models, where the CFG model has a tail in small and large values of $\omega$. At the peak region of the differential cross section, the CFG model prediction is smaller than the RFG model prediction. The agreement with the data varies depending on the electron energy and scattering angle (see Figs. \ref{Fig:480_MeV_60z}-\ref{Fig:240_MeV}), but the CFG model does not provide a systematic description of the data in all kinematic regimes considered. We also found that the differences between BBBA \cite{Bradford:2006yz} and BHLT parametrizations of the electromagnetic form factors were small compared to the differences between the nuclear models; see Fig. \ref{Fig:electron_FF_comparison}.

In Sec. \ref{sec:neutrino} we compared the predictions of the RFG and CFG models to neutrino-carbon scattering. The nucleon axial form factor plays an important role in this interaction. We used two ``extremes" of possible parametrizations of $F_A$, MBGH \cite{Meyer:2016oeg} and Mainz22 \cite{Djukanovic:2022wru}, with other parametrizations lying in between them.

First, we considered the hypothetical case of the fixed neutrino energy, $1$ GeV, around the peak of the MiniBooNE neutrino energy flux. At the tail regions, we could clearly distinguish between the RFG and CFG models, independent of the axial form factor parametrization. At the peak region, we could distinguish between RFG and CFG model predictions only for the Mainz22 parametrization, and not for the MBGH parametrization; see Figs. \ref{Fig:Fixed_Energy0.85} and \ref{Fig:Fixed_Energy0.75}.

Next, we compared to MiniBooNE data the flux-averaged neutrino cross section. Unlike the scattering of neutrinos with fixed energy, we could not distinguish between the RFG and CFG models using either MBGH or Mainz22. This is a somewhat disappointing result. It would be interesting to see if this phenomenon still persists in less inclusive observable, e.g., semi-inclusive neutrino scattering. For both models, Mainz22 fits the data much better than MBGH. This is to be expected, since Mainz22 $F_A$ largely overlaps with the dipole $F_A$ with $m_A^{\rm dipole}$ extracted by MiniBooNE \cite{MiniBooNE:2010bsu} assuming the RFG model, without the addition of multinucleon processes \cite{Martini:2009, Martini:2010, Benhar:2010, Barbaro:2011, Amaro:2011, Nieves:2011, Nieves:2011b, Bodek:2011, Meucci:2011}; see Fig. \ref{Fig:FA_Cmprsn}. Finally, we used the larger set of $F_A$ parametrizations that lay between MBGH or Mainz22 together with the RFG and CFG nuclear models. We found an almost continuous spread in the predictions generated using  these $F_A$ parametrizations for both RFG and CFG nuclear models. This highlights the need to get a smaller and more consistent uncertainty for $F_A$.

We hope that the analytic implementation of the CFG model we presented can be easily included in neutrino event generators. The axial form factor uncertainty for the MiniBooNE kinematics should be relevant to other experiments on the Booster Neutrino Beam, such as MicroBooNE \cite{Abratenko:2023} and SBND \cite{Antonello:2015}. It would be interesting to study the axial form factor uncertainty for these experiments, too. The analytic implementation can be adapted in the future to less-inclusive observables. For example, one can consider the kinematics of the final-state nucleon by not integrating over the final-state nucleon in Eq. (\ref{eq:General_W_H_relation}). In the future, it would be also interesting to combine the CFG model with other effects, such as FSIs, for example, as was done in Ref. \cite{Ankowski:2014yfa}.

\vskip 0.2in
\noindent
{\bf ACKNOWLEDGMENTS}
\vskip 0.1in
\noindent
We thank Jameson Tockstein, Joseph Wieske, and Malik Swain for collaboration in earlier stages of the project. We also thank Adi Ashkenazi, Or Hen, Kevin McFarland, and Eliezer Piasetzky for useful discussions. This work was supported by the U.S. Department of Energy Grant No. DE-SC0007983 (S.C., G.P.) and the National Science Foundation under Grant No.~PHY-2310627 (B.B.). 

\begin{appendix}
\section{Appendix: Values of parameters}
We present in Table \ref{table_parameters} the values of the parameters used in this paper and the reference to each value.
\begin{table}[h!]
\caption{Values of parameters \label{table_parameters}}
\begin{center}
\begin{tabular}{c|c|c|c} 
Name & Parameter & Value and unit& Reference \\
\hline
EM fine structure constant& $\alpha$ & $1/137.036$ & \cite{ParticleDataGroup:2022pth} \\ 
 Proton mass & $m_{p}$ & 938.272 MeV & \cite{ParticleDataGroup:2022pth} \\
Neutron mass  & $m_{n}$ & 939.565 MeV & \cite{ParticleDataGroup:2022pth} \\
 Proton magnetic moment & $\mu_p$ & 2.79285 & \cite{ParticleDataGroup:2022pth} \\ 
 Neutron magnetic moment& $\mu_n$ & $-1.91304$ & \cite{ParticleDataGroup:2022pth} \\ 
 Vector mass& $m_V$ & 843 MeV & \cite{Galster:1971kv} \\
 Pion mass & $m_\pi$ & 139.57 MeV & \cite{ParticleDataGroup:2022pth} \\
Carbon binding energy & $\epsilon_b$ & 25 MeV & \cite{Moniz:1971mt} \\ 
Carbon Fermi momentum  & $p_F$ & 220 MeV & \cite{MiniBooNE:2010bsu} \\
CFG model parameter  &$\lambda$ & $2.75\pm 0.25$ & \cite{Hen:2014yfa} \\
CFG model parameter  & $c_0$ & $4.16\pm 0.95$ & \cite{Hen:2014yfa} \\
 \hline
\end{tabular}
\end{center}
\end{table}

\section{Appendix: Expressions for $a$ and $b$ functions} \label{app:ab_functions}
Following Ref. \cite{Bhattacharya:2011ah}, we find the expressions for $a_i$s
to be as follows:
\begin{eqnarray}
a_1 &=& \int d^3{\bf p} f({\bf p}, q^0, {\bf q}) ~,~~ \\
a_2 &=& \int d^3{\bf p} f({\bf p}, q^0, {\bf q})~\dfrac{|{\bf p}|^2}{m_N^2} ~=~ \int d^3{\bf p}
f({\bf p}, q^0, {\bf q})~\dfrac{E_{\bf p}^2 - m^2_N}{m_N^2} ~,~~ \\
a_3 &=& \int d^3{\bf p} f({\bf p}, q^0, {\bf q})~\dfrac{p^2_z}{m_N^2} ~=~ \int d^3{\bf p}
f({\bf p}, q^0, {\bf q})~\cos^2\theta_{{\bf p}{\bf q}}~\dfrac{E_{\bf p}^2 - m^2_N}{m_N^2} ~,~~ \\
a_4 &=& \int d^3{\bf p} f({\bf p}, q^0, {\bf q})~\dfrac{\epsilon_{\bf p}^2}{m_N^2} ~=~ \int d^3{\bf p}
f({\bf p}, q^0, {\bf q})~\dfrac{(E_{\bf p} - \epsilon_b)^2}{m_N^2} ~,~~ \\
a_5 &=& \int d^3{\bf p} f({\bf p}, q^0, {\bf q})~\dfrac{\epsilon_{\bf p} p_z}{m_N^2} ~=~ \int d^3
{\bf p} f({\bf p}, q^0, {\bf q})~\cos\theta_{{\bf p}{\bf q}}~\dfrac{(E_{\bf p} - \epsilon_b)|{\bf p}|}{m_N^2}
~,~~ \\
a_6 &=& \int d^3{\bf p} f({\bf p}, q^0, {\bf q})~\dfrac{p_z}{m_N} ~=~ \int d^3{\bf p} f({\bf p},
q^0, {\bf q})~\cos\theta_{{\bf p}{\bf q}}~\dfrac{|{\bf p}|}{m_N} ~,~~ \\
a_7 &=& \int d^3{\bf p} f({\bf p}, q^0, {\bf q})~\dfrac{\epsilon_{\bf p}}{m_N} ~=~ \int d^3{\bf p}
f({\bf p}, q^0, {\bf q})~\dfrac{(E_{\bf p} - \epsilon_b)}{m_N} ~,~~
\end{eqnarray}
where $\epsilon_{\bf p} ~=~ E_{\bf p} - \epsilon_b~,~\epsilon'_{{\bf p}'} ~=~ E_{{\bf p}'}~,~ E_{{\bf p}}^2 ~=~
{\bf p}^2 + m_N^2 ~.~$ The integrals  have the generic form %
\begin{eqnarray}
\int d^3{\bf p} f({\bf p}, q^0, {\bf q}) X({\bf p}) &=& \dfrac{m_T V}{4\pi^2}\int d^3{\bf p}~n_i
({\bf p})(1 - n_f({\bf p} + {\bf q}))\dfrac{\delta (\epsilon_{\bf p} - \epsilon'_{{\bf p} + {\bf q}} + q^0)}
{\epsilon_{{\bf p}}\epsilon'_{{\bf p} + {\bf q}}}X({\bf p}) ~,~~~~ \nonumber\\
&=& \dfrac{m_T V}{4\pi^2}\int \dfrac{d^3{\bf p}}{(E_{\bf p} - \epsilon_b)|{\bf p}||{\bf q}|}~n_i
({\bf p})(1 - n_f({\bf p} + {\bf q})) ~~~ \nonumber\\
&&\hskip1in\times~\delta \left(\cos\theta_{{\bf p}{\bf q}} - \dfrac{\omega^2_{\rm eff} - |{\bf q}|^2 + 2 \omega_
{\rm eff} E_{{\bf p}}}{2|{\bf p}||{\bf q}|}\right)X({\bf p}) ~,~~
\end{eqnarray}
where $X({\bf p})$ is a function of $\bm p$ and we have used
\begin{eqnarray}
\delta (\epsilon_{\bf p} - \epsilon'_{{\bf p} + {\bf q}} + q^0) ~=~ \dfrac{E_{{\bf p} + {\bf q}}}{|{\bf p}||{\bf q}|}~
\delta \left(\cos\theta_{{\bf p}{\bf q}} - \dfrac{\omega^2_{\rm eff} - |{\bf q}|^2 + 2 \omega_{\rm eff} E_{{\bf p}}}
{2|{\bf p}||{\bf q}|}\right) ~.~~
\end{eqnarray}
The above delta function also enforces the condition $E_{{\bf p} + {\bf q}} = E_{\bf p} +
\omega_{\rm eff}$. For the CFG model, $n_{i,f}$ depends on either a constant or $|\bm p+\bm q|^{-4}$ or $|\bm p|^{-4}$. The integration  over the delta function simply enforces $|{\bf q}|^2 + 2{\bf p}\cdot{\bf q} = \omega_{\rm eff}^2 + 2 \omega_{\rm eff} E_{\bf p}$ giving  $|\bm p+\bm q|^{-4}=\left[(E_{\bf p} +\omega_{\mbox{\scriptsize eff}})^2-m_N^2\right]^{-2}$ and  $|\bm p|^{-4}=\left(E_{{\bf p}}^2 - m^2_N\right)^{-2}$. 

Let $B(E_{\bf p})$ be the expression for $n_i({\bf p})(1 - n_f({\bf p} + {\bf q}))$ after the integration over the delta function.  For each of the possible six transitions, we have

\begin{eqnarray}\label{eq:Bfunctions}
B({\rm I} \to {\rm III}) &=& \alpha_0\left(1 - \frac{\alpha_1}{[(E_{{\bf p}} + \weff)^2 - m^2_N]^2}\right)~,~~ \nonumber\\
B({\rm I} \to {\rm IV}) &=& \alpha_0 ~,~~  \nonumber\\
B({\rm I} \to {\rm V}) &=& \alpha_0(1 - \alpha_0) ~,~~  \nonumber\\
B({\rm II} \to {\rm III}) &=& \frac{\alpha_1}{(E_{{\bf p}}^2 - m^2_N)^2} \left(1 - \frac{\alpha_1}{[(E_{{\bf p}} + \weff]^2 - m^2_N)^2}\right)~,~~  \nonumber\\
B({\rm II} \to {\rm IV}) &=& \frac{\alpha_1}{(E_{{\bf p}}^2 - m^2_N)^2} ~,~~  \nonumber\\
B({\rm II} \to {\rm V}) &=& \frac{\alpha_1(1 - \alpha_0)}{(E_{{\bf p}}^2 - m^2_N)^2}~.~~
\end{eqnarray}
These six transitions have only four independent functional forms.  

Defining, as in the RFG model, $c = -\omega_{\rm eff}/|{\bf q}|$ and $d = -(\omega_{\rm eff}^2 - |{\bf q}|^2)/(2|{\bf q}|
m_N)$, and integrating over the delta function gives 
\begin{eqnarray}
\cos\theta_{{\bf p}{\bf q}}&\to& \dfrac{\omega^2_{\rm eff} - |{\bf q}|^2 + 2 \omega_{\rm eff} E_{{\bf p}}}{2|{\bf p}||{\bf q}|} =
-\dfrac{m_N d + E_{\bf p} c}{|{\bf p}|} ~,~~ \nonumber\\
\cos^2\theta_{{\bf p}{\bf q}}&\to&\left(\dfrac{\omega^2_{\rm eff} - |{\bf q}|^2 + 2 \omega_{\rm eff} E_{{\bf p}}}{2|{\bf p}||{\bf q}|}\right)^2 =
\left(\dfrac{m_N d + E_{\bf p} c}{|{\bf p}|}\right)^2=\dfrac{E_{\bf p}^2 c^2 + 2 m_N E_{\bf p} c d + m_N^2 d^2}{E_{\bf p}^2 - m_N^2}.~ ~~
\end{eqnarray}

All together, we have 
\begin{eqnarray}\label{eq:ai_final}
a_1 &=& \dfrac{m_T V}{2\pi|{\bf q}|}\int \dfrac{E_{\bf p} dE_{\bf p} }{(E_{\bf p} - \epsilon_b)}
~ B(E_{\bf p}) ~,~~ \nonumber\\
a_2 &=& \dfrac{m_T V}{2\pi|{\bf q}|}\int \dfrac{E_{\bf p} dE_{\bf p} }{(E_{\bf p} - \epsilon_b)}
\left(\dfrac{E_{\bf p}^2}{m_N^2} - 1\right)~ B(E_{\bf p}) ~,~~\nonumber\\
a_3 &=& \dfrac{m_T V}{2\pi|{\bf q}|}\int \dfrac{E_{\bf p} dE_{\bf p} }{(E_{\bf p} - \epsilon_b)}
\left(c^2 \dfrac{E_{\bf p}^2}{m_N^2} + 2 c d \dfrac{E_{\bf p}}{m_N} + d^2\right)~ B(E_{\bf p})
~,~~ \nonumber\\
a_4 &=& \dfrac{m_T V}{2\pi|{\bf q}|}\int \dfrac{E_{\bf p} dE_{\bf p} }{(E_{\bf p} - \epsilon_b)}
\left(\dfrac{E_{\bf p}^2}{m_N^2} - 2 \dfrac{\epsilon_b}{m_N}\dfrac{E_{\bf p}}{m_N} + \dfrac{
\epsilon_b^2}{m_N^2}\right)~ B(E_{\bf p}) ~,~~\nonumber\\
a_5 &=& \dfrac{m_T V}{2\pi|{\bf q}|}\int \dfrac{E_{\bf p} dE_{\bf p} }{(E_{\bf p} - \epsilon_b)}
\left(-~\dfrac{E_{\bf p}^2}{m_N^2} c + \left(\dfrac{\epsilon_b}{m_N} c - d\right)\dfrac{E_{\bf p}}{m_N}
+ \dfrac{\epsilon_b d}{m_N}\right)~ B(E_{\bf p}) ~,~~ \nonumber\\
a_6 &=& \dfrac{m_T V}{2\pi|{\bf q}|}\int \dfrac{E_{\bf p} dE_{\bf p} }{(E_{\bf p} - \epsilon_b)}
\left(-~\dfrac{E_{\bf p}}{m_N} c - d\right)~ B(E_{\bf p}) ~,~~ \nonumber\\
a_7 &=& \dfrac{m_T V}{2\pi|{\bf q}|}\int \dfrac{E_{\bf p} dE_{\bf p} }{(E_{\bf p} - \epsilon_b)}
\left(\dfrac{E_{\bf p}}{m_N} - \dfrac{\epsilon_b}{m_N}\right)~ B(E_{\bf p}) ~.~~ 
\end{eqnarray}

These integrals can be expressed in terms of four sets of three master integrals:
\begin{eqnarray}\label{eq:bij}
b_1^j  &=& \dfrac{m_T V}{2\pi|{\bf q}|} \int\dfrac{E_{\bf p} dE_{\bf p}}{(E_{\bf p} - \epsilon_b)}\left(\dfrac{E_{\bf p}}{m_N}\right)^j ~,~~ \nonumber\\
&=& \dfrac{m_T V}{2\pi|{\bf q}|m_N^j} \int\dfrac{E_{\bf p}^{j+1}dE_{\bf p}}{(E_{\bf p} - \epsilon_b)} ~,~~ \nonumber\\
b_2^j  &=& \dfrac{m_T V}{2\pi|{\bf q}|} \int\dfrac{E_{\bf p} dE_{\bf p}}{(E_{\bf p} - \epsilon_b)(E^2_{\bf p} - m^2_N)^2}\left(\dfrac{E_{\bf p}}{m_N}\right)^j ~,~~ \nonumber\\
&=& \dfrac{m_T V}{2\pi|{\bf q}|m_N^j} \int\dfrac{E_{\bf p}^{j+1}dE_{\bf p}}{(E_{\bf p} - \epsilon_b)(E_{\bf p}^2 - m^2_N)^2} ~,~~ \nonumber\\
b_3^j  &=& \dfrac{m_T V}{2\pi|{\bf q}|} \int\dfrac{E_{\bf p} dE_{\bf p}}{(E_{\bf p} - \epsilon_b)((E_{\bf p}+\weff)^2 - m^2_N)^2}\left(\dfrac{E_{\bf p}}{m_N}\right)^j ~,~~ \nonumber\\
&=& \dfrac{m_T V}{2\pi|{\bf q}|m_N^j} \int\dfrac{E_{\bf p}^{j+1}dE_{\bf p}}{(E_{\bf p} - \epsilon_b)((E_{\bf p} + \weff)^2 - m^2_N)^2} ~,~~ \nonumber\\
b_4^j &=& \dfrac{m_T V}{2\pi|{\bf q}|} \int\dfrac{E_{\bf p} dE_{\bf p}}{(E_{\bf p} - \epsilon_b)(E_{\bf p}^2 - m^2_N)^2((E_{\bf p}+\weff)^2 - m^2_N)^2}\left(\dfrac{E_{\bf p}}{m_N}\right)^j ~,~~ \nonumber\\
&=& \dfrac{m_T V}{2\pi|{\bf q}|m_N^j} \int\dfrac{E_{\bf p}^{j+1}dE_{\bf p}}{(E_{\bf p} - \epsilon_b)(E_{\bf p}^2 - m^2_N)^2((E_{\bf p} + \weff)^2 - m^2_N)^2},~~
\end{eqnarray} 
where $j=0,1,2$. Note that $b_1^j$ is equal to $b_j$ as defined in Ref.\cite{Bhattacharya:2011ah} for the RFG case.\mbox{}\\ As an example, the integral $a_3$ for the six possible transitions are written in terms of $b_{1,2,3,4}^j$ as
\begin{eqnarray}
a_3({\rm I} \to {\rm III}) &=& \alpha_0\Big(c^2b_1^2+2cdb_1^1+d^2b_1^0 - \alpha_1 (c^2b_3^2 +2cdb_3^1+d^2b_3^0) \Big)~,~~ \nonumber\\
a_3({\rm I} \to {\rm IV}) &=& \alpha_0\Big(c^2b_1^2+2cdb_1^1+d^2b_1^0\Big)~,~~ \nonumber\\
a_3({\rm I} \to {\rm V}) &=& \alpha_0(1-\alpha_1)\Big(c^2b_1^2+2cdb_1^1+d^2b_1^0\Big)~,~~ \nonumber\\
a_3({\rm II} \to {\rm III}) &=& \alpha_1\Big(c^2b_2^2+2cdb_2^1+d^2b_2^0 - \alpha_1 (c^2b_4^2 +2cdb_4^1+d^2b_4^0) \Big)~,~~ \nonumber\\
a_3({\rm II} \to {\rm IV}) &=& \alpha_1\Big(c^2b_2^2+2cdb_2^1+d^2b_2^0\Big)~,~~ \nonumber\\
a_3({\rm II} \to {\rm V}) &=& \alpha_1(1-\alpha_0)\Big(c^2b_2^2+2cdb_2^1+d^2b_2^0\Big)~.~~
\end{eqnarray}
Similar expressions hold for other $a_i$.
Similarly, as an example, one of the coefficients $W_1$ of the nuclear tensor which are functions of $b_{1,2,3,4}^j$ and the form factor for transition ${\rm I} \to {\rm V}$ is -
\begin{eqnarray*}
W_1^{{\rm I} \to {\rm V}}&=&\frac{3 \pi A^2 \alpha_0 (1-\alpha_0)(m_N-\epsilon_b)}{32 m_N^2 p_F^3 \bm{q}^3} \Bigg[\Big\{\left(\bm{q}^2-(\epsilon_b-\omega)^2\right)^2 b_1^0 + 4m_N \left(\bm{q}^2-(\epsilon_b - \omega)^2\right)(\epsilon_b-\omega)b_1^1 \\&+& 4m_N^2 \bm{q}^2 \left(b_1^0-b_1^2\right)+4m_N^2 (\epsilon_b-\omega)^2 b_1^2\Big\}\Big\{2q^2 F_2(q^2)^2 -8m_N^2 \left(F_1(q^2)^2+F_A(q^2)^2\right)\Big\}\\&+& 8m_N^2 \bm{q}^2 b_1^0 \left\{8m_N^2 F_A(q^2)^2-2q^2 \left(\left(F_1(q^2)+F_2(q^2)\right)^2+F_A(q^2)^2\right)\right\}\Bigg]
\end{eqnarray*}

\section{Appendix: Form factors parametrizations}\label{app:form_factors}
We list here the form factors parametrizations considered in this paper. We refer to the original papers for the values of the parameters used for each parametrization. The error bars are calculated based on the information given in these original papers.
\subsection{Vector form factor} 
\subsubsection{BBBA parametrization}
The BBBA parametrization \cite{Bradford:2006yz} uses the functional form of 
\begin{eqnarray}
 G_{E/M}^{p,n}=\frac{\sum_{k=0}^2 a_k \tau^k}{1 + \sum_{k=1}^4 b_k \tau^k}\,.
\end{eqnarray}
The parameters $a_k$ and $b_k$ are listed in Ref. \cite{Bradford:2006yz}.
\begin{figure}[h!]
\begin{center}
\includegraphics[scale=0.9]{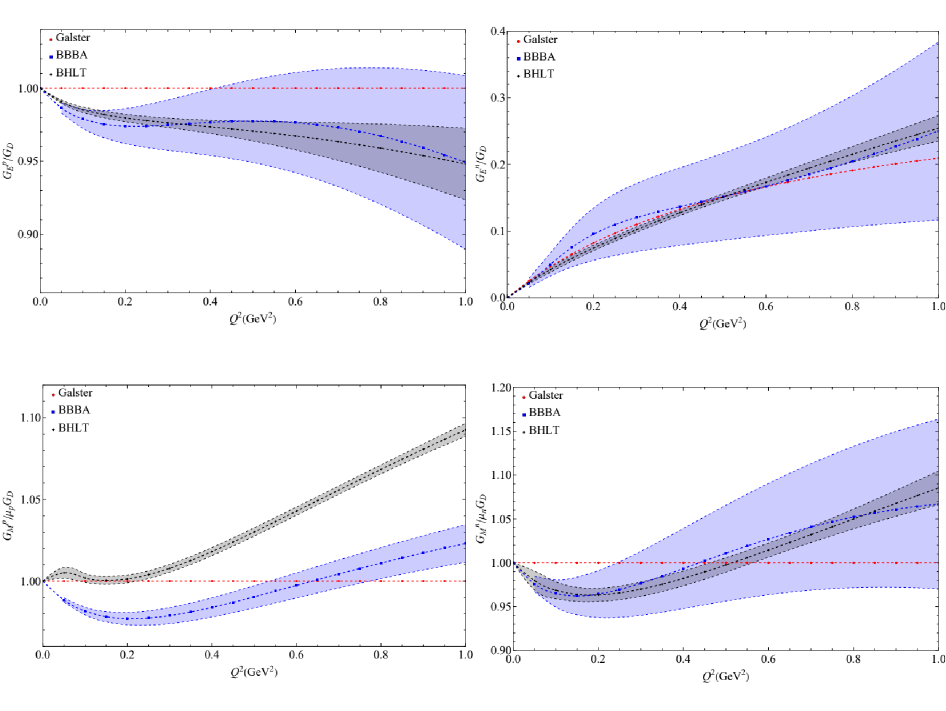}
\caption{\label{Fig:EM_FF_Cmprsn} Comparison between the different electromagnetic form factor parametrizations shown for $G^p_E$ and $G^n_E$ (top) $G^p_M$ and $G^n_M$ (bottom). The red circle, blue square, and gray diamond data points represent the Galster, BBBA, and BHLT Parametrizations, respectively.}
\end{center}
\end{figure}
\subsubsection{BHLT parametrization}
The form factors in the BHLT parametrization \cite{Borah:2020gte} were determined from a global fit to electron scattering data and precise charge radius measurements. The form factors are expressed as a convergent expansion in $z(q^2)$,
\begin{eqnarray}
    G_E^{p,n} &=& \sum_{k=0}^{k_{max}} a_k z(q^2)^k, \nonumber\\
    G_M^{p,n} &=& G_M^{p,n} (0)\sum_{k=0}^{k_{max}} b_k z(q^2)^k,\nonumber \\
    z(q^2) &=& \frac{\sqrt{t_{cut} -q^2}-\sqrt{t_{cut} - t_0}}{\sqrt{t_{cut} -q^2}+\sqrt{t_{cut} - t_0}},
\end{eqnarray}
where, $t_{cut} = 4 m_\pi^2$ and $t_0 = -0.21\,$GeV$^2$. The form factor coefficients for the proton and neutron $a_1$ to $a_4$, and the corresponding covariant matrix for the uncertainty are provided in the supplementary material of Ref. \cite{Borah:2020gte}.\\  
To calculate the five unknown coefficients, namely $a_0$ and $a_5-a_8$, the normalization $G(0)$ ($G^p_E(0) = 1$, $G^n_E(0) = 0$ , $G^p_M(0) = \mu_p$, and $G^n_M(0) = \mu_n$) and the four sum rules in Eq. (6) from Ref. \cite{Borah:2020gte} are used:
\begin{eqnarray}\label{eq:sumrule}
    G(0) - \sum_{k=0}^8 a_k z(0)^k = 0, \quad \sum_{k=0}^8 a_k = 0, \quad \sum_{k=1}^8 k a_k = 0  \nonumber\\  
       \sum_{k=2}^8 k (k-1) a_k = 0 , \quad \sum_{k=3}^8 k (k-1) (k-2) a_k = 0\,.
\end{eqnarray}
In Fig. \ref{Fig:EM_FF_Cmprsn} we present a comparison of BBBA and BHLT parametrizations, as well as the historical Galster \cite{Galster:1971kv} parametrization that is sometimes used, e.g., in Ref. \cite{Sobczyk:2019yiy}.

\subsection{Axial form factor} With the exception of the historical dipole model, given in Sec. \ref{sec:neutrino},  axial form factor parametrizations are extracted using $z$-expansion, where the axial form factor $F_A$ is given by
\begin{equation}\label{eq:FA}
    F_A(q^2) = \sum_{k=0}^{k_{max}} a_k z(q^2)^k,\quad \mbox{where} \quad z(q^2) = \frac{\sqrt{t_{cut} -q^2}-\sqrt{t_{cut} - t_0}}{\sqrt{t_{cut} -q^2}+\sqrt{t_{cut} - t_0}}.
\end{equation}
\subsubsection{MBGH parametrization}
The axial  nucleon form factor in the MBGH parametrization \cite{Meyer:2016oeg} was determined from charged-current neutrino-deuterium scattering data. The $z$-expansion from Eq. (\ref{eq:FA}) is used to calculate the form factor for $k_{max} = 8$ with, $t_{cut} = 9 m_\pi^2$ ($m_{\pi} = 140$ MeV), and $t_0 = -0.28\,$GeV$^2$. The coefficients for $k = 1$ to $4$  and the covariant matrix are listed in Ref. \cite{Meyer:2016oeg}. The five remaining coefficients are determined from normalization, $F_A(0) = g_A = -1.2723$, and the sum rule constraints as in Eq. (\ref{eq:sumrule}). 
\subsubsection{NME 22}
The axial  form factor by the Nuclear Matrix Element Collaboration \cite{Park:2021ypf} is extracted from the results for the axial current between ground-state nucleons. The parametrization in the $\{4^{N\pi},2^{sim},\hat{z}^2\}$ fit is obtained using the $z$-expansion from Eq. (\ref{eq:FA}) for $k_{max} = 2$ with $t_{cut} = 9 m_\pi^2$ ($m_{\pi} = 135$ MeV), and $t_0 = -0.5\,$GeV$^2$.
The three parameters and the covariant matrix can be found in Ref. \cite{Park:2021ypf}.
\subsubsection{Mainz22}
The $z$-expansion coefficients were extracted directly from lattice correlators by the Mainz group. Using the $z$-expansion from Eq. (\ref{eq:FA}) for $k_{max} = 2$ with, $t_{cut} = 9 m_\pi^2$ ($m_{\pi} = 135$ MeV), and $t_0 = 0\,$GeV$^2$ the axial  form factor was obtained.
The coefficients for $k = 0$ to $2$ and the covariant matrix for the three coefficients are listed in Ref. \cite{Djukanovic:2022wru}.
\begin{figure}[H]
\begin{center}
\includegraphics[scale=0.35]{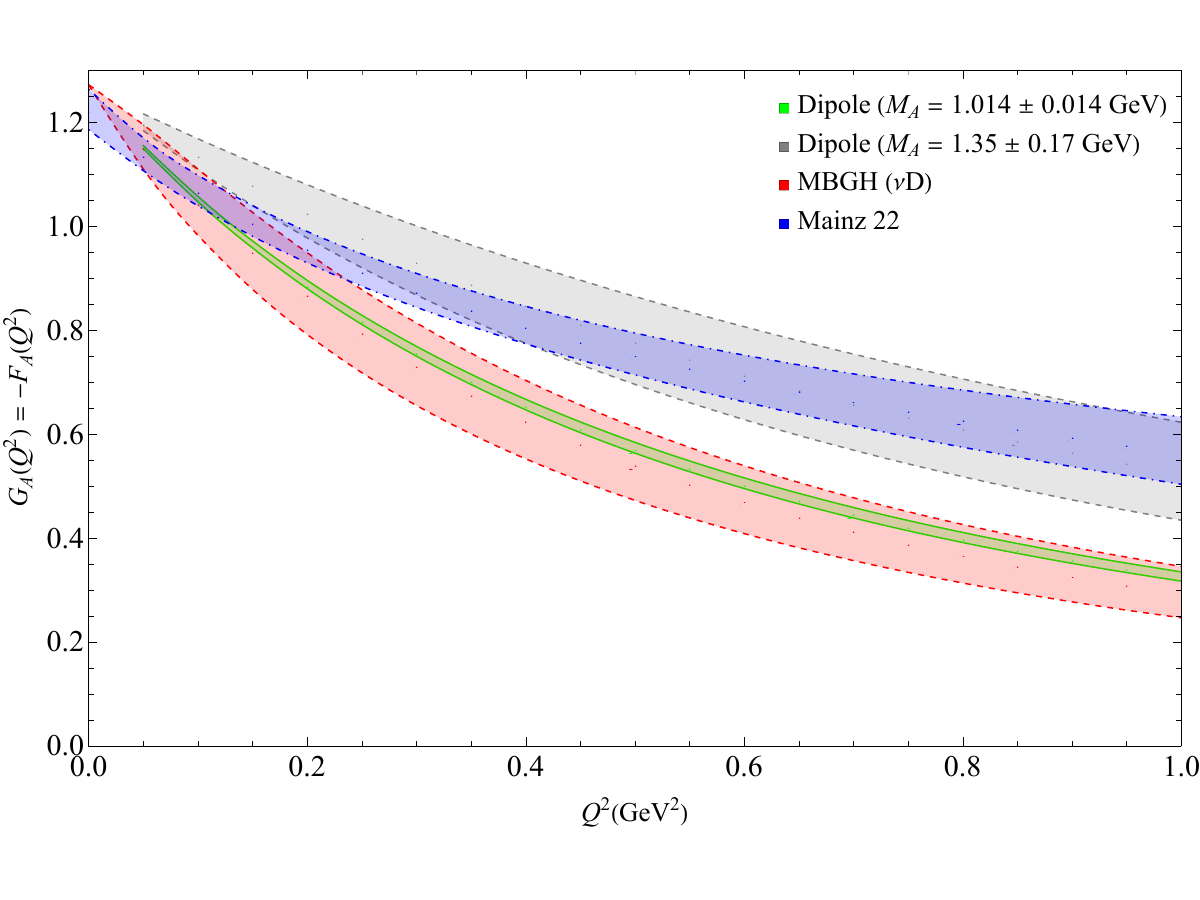}
\hspace{1cm}
\includegraphics[scale=0.35]{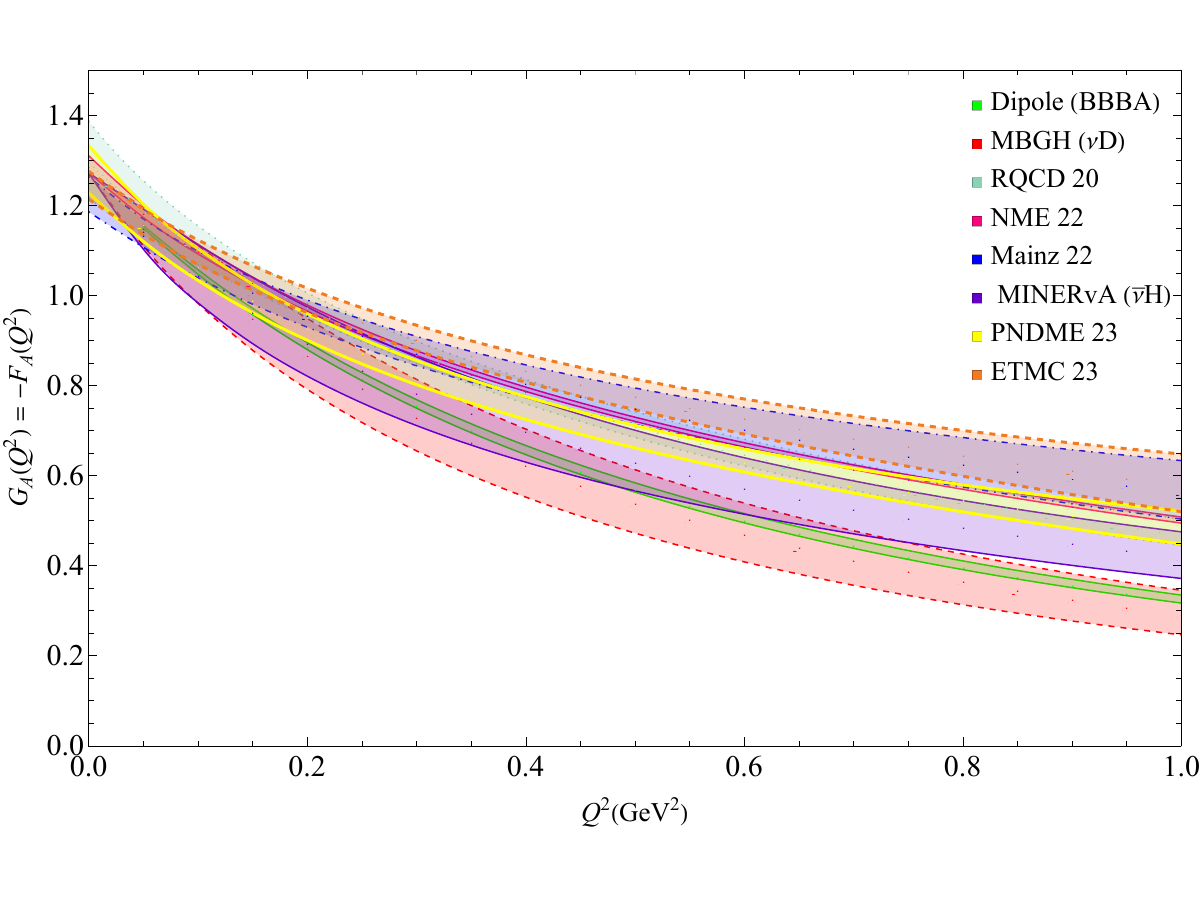}
\caption{\label{Fig:FA_Cmprsn} Comparison between the different axial  form factor parametrizations implemented for neutrino scattering.  Left: comparison between the models Mainz22 and MBGH which span across the two dipole models with $M_A = 1.35\pm0.17$ GeV \cite{MiniBooNE:2010bsu} and $M_A = 1.014\pm0.014$ GeV \cite{Bodek:2007ym}, respectively. Right: comparison between dipole (BBBA) using $M_A = 1.014\pm0.014$ GeV, MBGH \cite{Meyer:2016oeg}, RQCD 20 \cite{RQCD:2019jai}, NME 22 \cite{Park:2021ypf}, Mainz22 \cite{Djukanovic:2022wru}, MINERvA \cite{MINERvA:2023avz}, PNDME 23 \cite{Jang:2023zts}, and ETMC 23 \cite{Alexandrou:2023qbg}.}
\end{center}
\end{figure}
\subsubsection{MINERvA}
Axial form factor is extracted from the $\bar\nu$-hydrogen scattering using the plastic scintillator target of the MINERvA experiment \cite{MINERvA:2023avz}. The $z$-expansion from Eq. (\ref{eq:FA})  is used to extract $F_A$ from the hydrogen cross section for $k_{max} = 8$, with $t_{cut} = 9 m_{\pi}^2$ ($m_\pi = 140$ MeV), and $t_0 = -0.75\,$GeV$^2$. The coefficients for $k = 0$ to $4$  and the covariant matrix are listed in Ref. \cite{MINERvA:2023avz}. The five remaining coefficients are determined from normalization ($F_A(0) = g_A = -1.2723$) and the sum rule constraints in Eq. (\ref{eq:sumrule}).
\subsubsection{PNDME 23}
 The coefficients in $F_A$ are extracted using thirteen 2+1+1 flavor highly improved staggered quark ensembles. Using $k_{max} = 2$ in Eq. (\ref{eq:FA}), with $t_{cut} = 9 m_{\pi}^2$ ($m_\pi = 135$ MeV), and $t_0 = -0.25\,$GeV$^2$ the axial form factor is obtained. The coefficients for $k = 0$ to $2$  and the covariant matrix are provided in Ref. \cite{Jang:2023zts}. 
\subsubsection{ETMC 23}
The axial form factor in the Extended Twisted Mass Collaboration \cite{Alexandrou:2023qbg} is evaluated using three $N_f = 2+1+1$ twisted mass fermion ensembles. The $z$-expansion from Eq. (\ref{eq:FA}) is used to extract $F_A$ from the hydrogen cross section, with $t_{cut} = 9 m_{\pi}^2$ ($m_\pi = 135$ MeV), and $t_0 = 0\,$GeV$^2$. The coefficients for $k = 0$ to $3$  and the covariant matrix are provided in Ref. \cite{Alexandrou:2023qbg}.
\subsubsection{RQCD 20}
The two- and three-point correlation functions are extracted using EFT methods in \cite{RQCD:2019jai}. They obtain fits and in turn extract ground-state form factors. The $z$-expansion formalism is used to extract $F_A$, with $t_{cut} = 9 m_{\pi}^2$ and $t_0 = -t_{cut}$. No error bars are given for the $z$-expansion coefficients in Ref. \cite{RQCD:2019jai}, only for the form factor itself. Thus, we plot the form factor in Fig. \ref{Fig:FA_Cmprsn}, but we do not use it in Figs. \ref{Fig:Theta_0.85_RFG_CFG} and \ref{Fig:Theta_0.75_RFG_CFG}.

\section{Appendix: More plots for electron scattering} \label{app:plots}
\begin{figure}[h!]
\begin{center}
\includegraphics[scale=1.1]{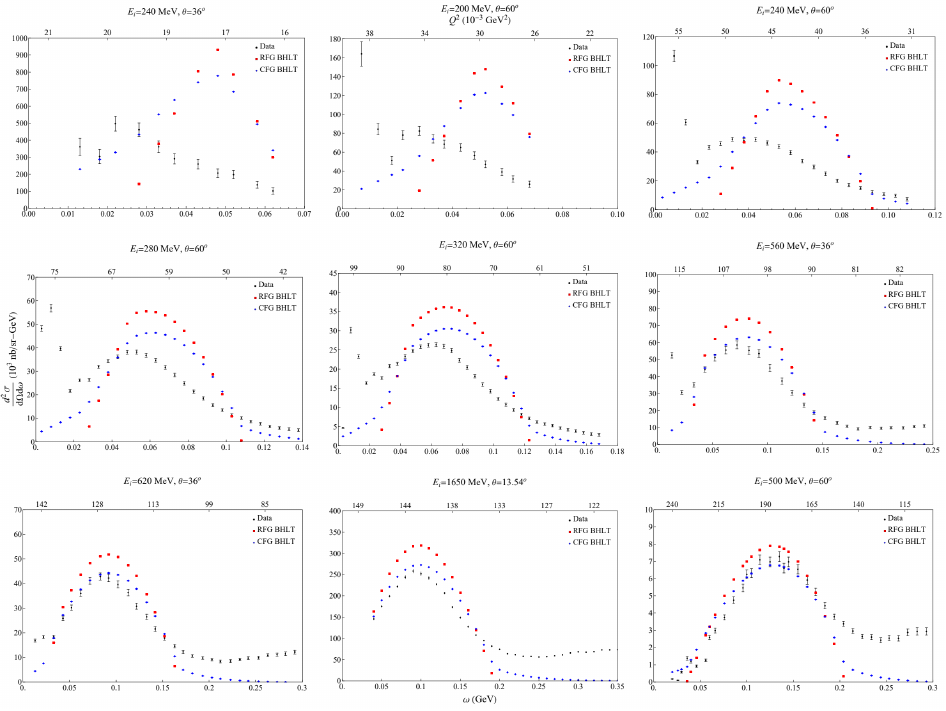}
\caption{\label{Fig:Electron_Scattrng_Collage} Comparison of double differential cross section using RFG (red squares) and CFG (blue diamonds) nuclear model with the carbon data for various incoming electron energy and scattering angles. For all plots the lower horizontal axis is $\bm\omega$ in GeV and the upper horizontal axis is $Q^2$ in units of $10^{-3}$ GeV.}
\end{center}
\end{figure}

In Fig. \ref{Fig:Electron_Scattrng_Collage} we present comparison of the RFG and CFG models to electron-carbon scattering data for more kinematical points. The points shown were also considered in Ref. \cite{Ankowski:2014yfa}.

\section{Appendix: Effects of varying the binding energy}\label{app:binding energy}

\begin{figure}[H]
\begin{center}
\includegraphics[scale=0.38]{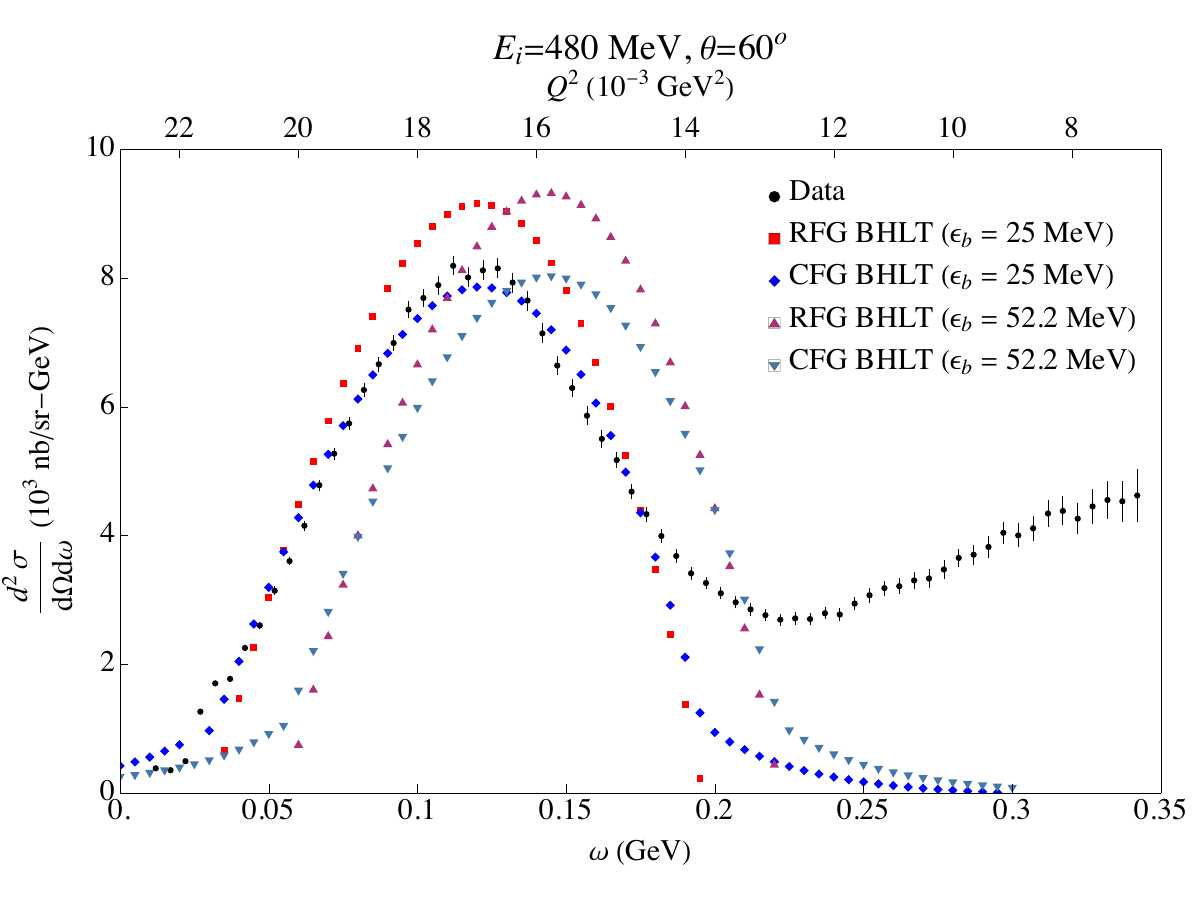}
\hspace{1cm}
\includegraphics[scale=0.38]{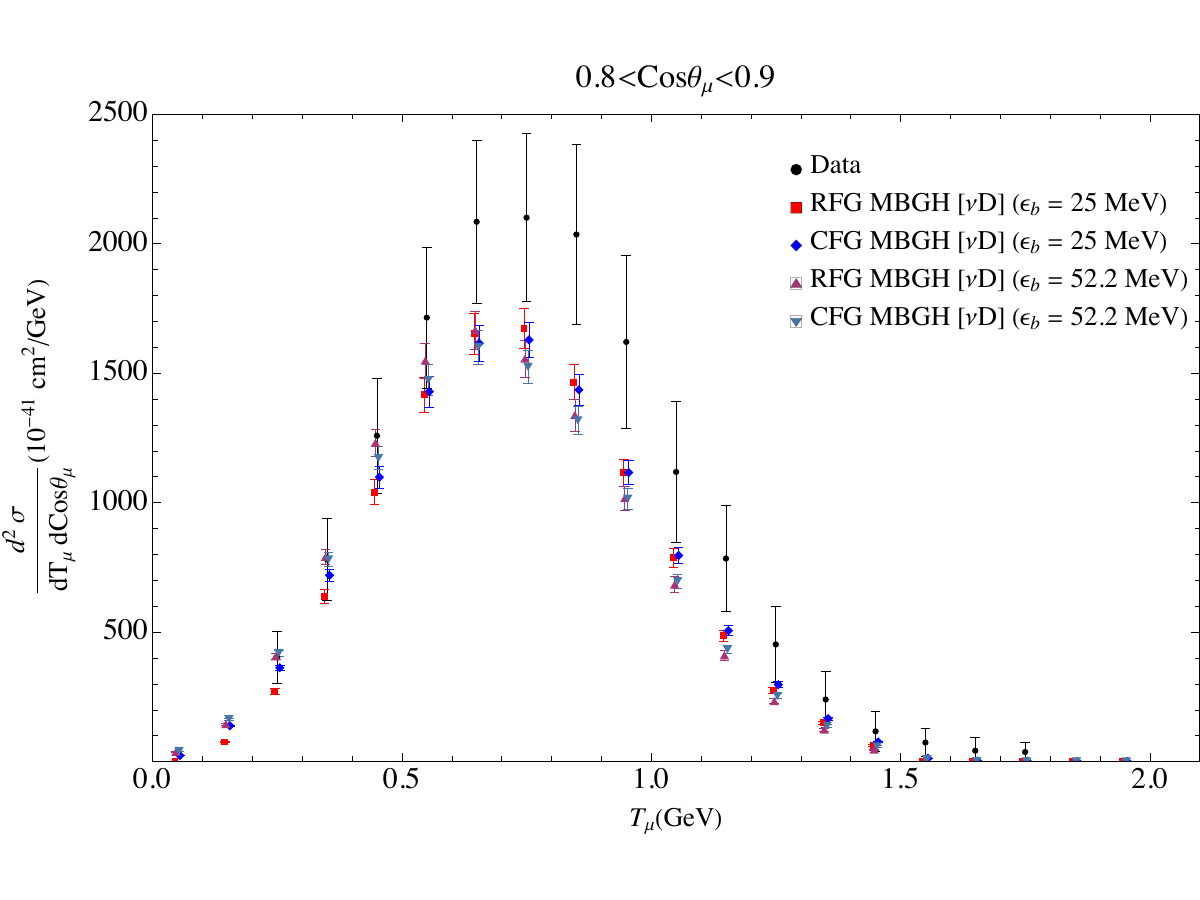}
\caption{\label{Fig:Binding_Energy_Cmprsn} Comparison of the RFG and CFG models for two different values of the binding energy, $\epsilon_b$. Left: differential cross section for electron scattering as a function of $\omega$, for fixed values of the incoming electron's energy, $E_i = 480$ MeV, and scattering angle, $\theta = 60^\circ$. Right: flux-averaged cross section for neutrino-nucleus scattering as a function of the lepton kinetic energy, plotted for $\cos\theta_\mu \in (0.8,0.9)$.}
\end{center}
\end{figure}

The binding energy, $\epsilon_b$, enters the calculations of the lepton-nucleus scattering cross section through Eq.~(\ref{eq:replacements}). In electron-nuclei scattering experiments reported in Ref. \cite{Moniz:1971mt} a binding energy of 25 MeV for $^{12}$C was found for the RFG model. In this paper, we use the same value for the CFG model. The GFMC approach finds a larger binding energy, 52.2 MeV \cite{Benhar:2012nj}. In Fig. \ref{Fig:Binding_Energy_Cmprsn}, we compare the results of using the RFG and CFG models for these two values of the binding energy ($\epsilon_b = 25$ MeV and $\epsilon_b = 52.2$ MeV). 

The left plot of Fig. \ref{Fig:Binding_Energy_Cmprsn} shows how the differential cross section changes with the binding energy for an incoming electron with $E_i = 480$ MeV and $\theta = 60^\circ$. Here, we use the BHLT parametrization for the nucleon's electric and magnetic form factors. For both the RFG and CFG models, we observe an overall shift in the distribution to higher values of $\omega$ as the binding energy is increased. This is to be expected, as the binding energy enters Eq.~(\ref{eq:replacements}) as a constant that offsets $\omega$.

The right plot of Fig. \ref{Fig:Binding_Energy_Cmprsn} shows how the flux-averaged cross section for neutrino-nucleus scattering changes with the binding energy for the bin with $0.8<\cos\theta_\mu<0.9$. Here, we use the MBGH parametrization for the axial form factor. We find that at lower values of $T_\mu$, a larger value of the binding energy leads to an increased cross section, whereas at higher values of $T_\mu$, a larger value of the binding energy leads to a decrease in the cross section. Still, for a fixed binding energy, we cannot distinguish between the RFG and CFG models.  

\section{Appendix: Spectral functions} \label{app:spctrl_functions}
To allow for an easier comparison with the literature, we present here the relation between the nuclear tensor $W^{\mu\nu}$ and the nucleon tensor $H^{\mu\nu}$, Eq. (\ref{eq:General_W_H_relation}) in the language of spectral functions for both the RFG and CFG models. We follow the notation of Ref. \cite{Sobczyk:2017mts}, which lists several such spectral functions. In terms of spectral functions, 
the nuclear tensor can be written as \cite{Sobczyk:2017mts} -
\begin{equation}\label{sf:nucl tensor}
    W^{\mu\nu} = \int \frac{d^3 p}{(2\pi)^3}\frac{3\, m_N^2}{4 \pi p_F^3}\int dE S^h(E,\bm{p})S^p(\omega-E,\bm{p+q})H^{\mu\nu},
\end{equation}
where $E$ is the removal energy, the superscript ``$h$" denotes the initial nucleon (``hole") and the superscript ``$p$" denotes the final nucleon (``particle").

For the RFG model, the spectral functions can be expressed as -
\begin{eqnarray}\label{sf:rfg}
    S^h_{RFG}(E,\bm{p}) &=& \frac{\kappa}{E_{\bm{p}}-\epsilon_b} \theta(p_F-|\bm{p}|) \delta(E+E_{\bm{p}}-\epsilon_b) \,,  \\
    S^p_{RFG}(\omega - E,\bm{p+q}) &=& \frac{\kappa}{E_{\bm{p+q}}} \theta(|\bm{p+q}|-p_F) \delta(\omega - E-E_{\bm{p+q}}) \,,
\end{eqnarray}
where $\kappa = 2\pi^2\sqrt{A\,m_T}/m_N$ and $E_{\bm{p}} \equiv \sqrt{m^2_N + |\bm{p}|^2}$. Analogous expressions can be found in Section III A of Ref. \cite{Sobczyk:2017mts}.

Using Eqs.~(\ref{eq:replacements}) and (\ref{eq:Bfunctions}), we express the spectral functions for the CFG model as -
\begin{eqnarray}
   S_{CFG}^h(E,\bm{p}) &=& \frac{\kappa}{E_{\bm{p}}-\epsilon_b}\Big[\alpha_0\,\theta( p_F - |\bm{p}|) \nonumber \\
   && \hspace{25truemm} +~\frac{\alpha_1}{|\bm{p}|^4}\theta( |\bm{p}|-p_F)\,\theta( \lambda \, p_F - |\bm{p}|)\Big]\delta(E+E_{\bm{p}}-\epsilon_b) \,, \\
    S_{CFG}^p(\omega - E,\bm{p+q}) &=& \frac{\kappa}{E_{\bm{p+q}}}\Big[(1-\alpha_0)\,\theta(p_F-|\bm{p+q}| )+\theta( |\bm{p+q}|-\lambda\, p_F) \nonumber\\
&& \hspace{-15truemm}+~\Big(1-\frac{\alpha_1}{|\bm{p+q}|^4}\Big)\,\theta( |\bm{p+q}|-p_F)\,\theta( \lambda\, p_F-|\bm{p+q}|) \Big]\delta(\omega - E-E_{\bm{p+q}})\,.~~~~~
\end{eqnarray}

\end{appendix}

\bibliography{CCQE}
\bibliographystyle{hunsrt}

\end{document}